\documentclass[useAMS,usenatbib]{mn2e}
\usepackage{lscape}
\usepackage{graphicx,times}
\usepackage{multirow}

\title[Physical properties, kinetics and mass function of 12 northern infrared dark clouds]
{Physical properties, kinematics and mass function of 12 northern
infrared dark clouds}

\author[Xiao-Lan Liu, Jun-Jie
Wang and Jin-Long Xu ]{Xiao-Lan Liu$^{1,3}$\thanks{E-mail:
liuxiaolan10@mails.gucas.ac.cn}, Jun-Jie Wang$^{1,2}$ and Jin-Long
Xu$^{1,2}$ \\
$^{1}$National Astronomical Observatories, Chinese Academy of
Science, Beijing 100012, China\\
$^{2}$NAOC-TU Joint Center for Astrophysics, Lhasa 850000, China \\
$^{3}$Graduate University of Chinese Academy of Sciences, Beijing,
100049, China}

\begin{document}
\date{Accepted 1988 December 15. Received 1988 December 14; in original form 1988 October 11}

\pagerange{\pageref{firstpage}--\pageref{lastpage}} \pubyear{2002}

\maketitle

\label{firstpage}

\begin{abstract}
The physical, chemical and kinetic characteristics of 12 northern infrared dark clouds (IRDCs) are systematic studied using the $\rm ^{13}CO $ (1-0) and $\rm
C^{18}O$ (1-0) lines, observed with the PMO 13.7 m radio telescope, the 1.1 mm Bolocam Galactic Plane Survey (BGPS) data and GLIMPSE Spitzer IRAC $\rm 8 \,\mu m$
data. The molecular lines emission and 1.1 mm continuum emission almost coincide in morphology for each IRDC and both are associated well with the IRDCs. 10 IRDCs
present the filamentary structure and substructures. Totally, 41 IRDC cores are identified and a statistic research for them shows that the northern IRDC cores have
a typical excitation temperature $8\sim10$ K, a integrated intensity ratio of $\rm ^{13}CO$ to $\rm C^{18}O$ $3\sim6$ and the column density $(1\sim6)\times
10^{22}\, \rm cm^{-2}$. About $57.5\%$ of the IRDC cores are gravitationally bound, which are more compact, warmer and denser. In addition, we study the mass
distribution functions of the whole IRDC cores as well as the gravitational bound cores, finding that they almost have the same power-law indexes. This indicates
that the evolution of the IRDC cores almost have no effect on the mass spectrum of the molecular cores and thus can be used to study the stellar initial mass
function. Moreover, three IRDC cores G24.00-3, G31.38-1 and G34.43-4 are detected to have large-scaled infall motions. Two different outflows are further found for
IRDC core G34.43-4 and one of them is in high collimation.
\end{abstract}


\section{Introduction}
Despite massive stars play an important role in determining the
galactic environment and enrichment, the formation and protostellar
evolution of massive stars (M$\rm >8\,M_\odot$) is still unclear. In
addition, observing the earliest evolutionary stages of massive
stars directly is a challenging task, due to their rarer numbers,
farther distances and forming predominantly in clusters. However, if
we can unveil the mystery of massive star formation, it will be of
primary importance to learn about the evolution of galaxy. Whatever,
investigations of high-mass star cores are important to answer a
central question in star formation: How do star formation processes
produce the stellar initial mass function (IMF)?

Infrared Dark Clouds (IRDCs) have been proposed to be the birthplace
of massive stars and their host clusters
\citep{egan98,carey98,rathborne06}. Infrared dark clouds (IRDCs) are
first discovered to be dark silhouettes against the bright galactic
mid-infrared background by the infrared Space Observatory (ISO)
surveys \citep{perault96} and the Midcourse Space Experiment (MSX)
\citep{carey98,egan98}. Previous molecular lines and dust
continuum researches suggested that IRDCs were the cold ($\rm T<25$
K), dense ($\sim10^5 \rm \,cm^{-3}$) and high column density ($\geq
10^{23}\, \rm cm^{-2}$) clouds, with a scale of $1\sim10$ pc and a
mass of $\rm10^2\sim10^5 \,M_\odot$ \citep{egan98,carey98,carey00,
rathborne06}. Furthermore, strong mm or sub-mm dust emissions have
been detected in the IRDC cores
\citep{beuther05,rathborne05,rathborne06,rathborne08}. All of these
properties imply that IRDCs are excellent candidates for hosting the
very early stages of massive star formation.

While \citet{simon06a} established a catalogue of MSX IRDCs containing 10,931 sources and
\citet{Peretto09} catalogued 11,303 IRDCs using the Spitzer GLIMPSE and MIPSGAL archive
data. The studies aiming at the distributions of the IRDCs demonstrated that most IRDCs
concentrated on the so-called Galactic molecular ring in the first galactic quadrant
\citep{simon06b} and the first-quadrant combining with the four-quadrant IRDCs closely
followed the Scutum-Centaurus arm \citep{jackson08}, implying that IRDCs had some
relationship with massive star formation. Five evolutional stages were proposed by
\citet{chambers09} through his investigation towards 190 cores of 38 IRDCs. Recently, a
great number of works focus on the chemistry in the IRDCs
\citep[e.g.][]{vasy11,sanhu12,liu13,Miettinen14}, in order to find out the perfect
molecular tracers to different evolutional phases, the elemental abundance, the chemical
clocks and all kinds of differences as well as relations between various molecules.

The molecule $\rm C^{18}O$ is relatively abundant and is likely to
be optically thin in the molecular clouds. Therefore, it can be used
to study the structures and densities of the clouds. In this paper,
we make mapping observations towards 12 northern IRDCs in $\rm
^{13}CO$ (1-0) and $\rm C^{18}O$ (1-0) lines to study the physical
and chemical properties, kinematics and core mass functions of the
northern IRDC cores. In the remaining part of this work, section 2
describes the source selection, the data achieve and the data reduction.
Section 3 details the direct results and section 4 gives a discussion of the star
formation activities in three IRDC cores and core mass functions.
Finally, we summarizes our conclusions in Section 5.

\section{Data Achieve}
\subsection{Source selection}
To make sure the selected sources are really IRDCs and cover more IRDCs in our
observed regions, we have observed all the IRDC sources from \citet{Parsons09} which are
accessible from the Purple Mountain Observatory. This IRDC sample is likely to contain a
number of clumps in the different evolutionary stages, ranging from IR-dark clumps to HII
regions with bright IR emission. This unbiased selection increases the credibility and
representativeness to explore the whole physical, chemical and dynamical properties of the
IRDCs in the northern sky, since currently we have no way to observe all the IRDCs in the
northern sky.
\subsection{Observation}
The observations towards the IRDCs in $\rm ^{12}CO$ (1-0), $\rm ^{13}CO$ (1-0) and
$\rm C^{18}O$ (1-0) lines were carried out with the Purple Mountain Observatory (PMO)
13.7 m radio telescope in May 2012. The new 9-beam array receiver system in
single-sideband (SSB) mode was used as front end. FFTS spectrometers were used as back
end, which had a total bandwidth of 1 GHz and 16384 channels, corresponding to a velocity
resolution of 0.16 km $\rm s^{-1}$ for $\rm^{12}CO$ (1-0) and 0.17 km $\rm s^{-1}$ for
$\rm^{13}CO$ (1-0) and $\rm C^{18}O$ (1-0). $\rm^{12}CO$ (1-0) was observed at upper
sideband, while $\rm^{13}CO$ (1-0) and $\rm C^{18}$O (1-0) were observed simultaneously
at lower sideband. The half-power beam width (HPBW) was $\sim 53''$ and the main beam
efficiency was $\sim 0.5$. The pointing accuracy of the telescope was better than $4''$.
The system noise temperature ($\rm T_{sys}$) in SSB mode varied between 150 K and 400 K.
The On-The-Fly (OTF) observing mode was applied. The antenna continuously scanned a
region of $20'\times20'$ with a scan speed of $\rm 30''\, s^{-1}$ for each IRDC. However,
the edges of the OTF maps were very noisy and thus only the central $6.5'\times6.5'$
regions were selected to be further analyzed. Since CO is the second abundant molecule in
the Milky Way galaxy, its spectra can be easily affected by the objects along the line of
sight, so were our $\rm^{12}CO$ (1-0) observation data. Therefore, we did not use the
$\rm^{12}CO$ (1-0) observation data. The rms noise level was 0.1-0.2 K for $\rm^{13}CO$
(1-0) and $\rm C^{18}O$ (1-0). The data were reduced by the software CLASS (Continuum and
Line Analysis Single-Disk Software) and GREG (Grenoble Graphic).

Of the IRDC regions observed, only 12 were detected with a signal to noise ratio of
3 or greater and only these well detected regions are discussed further. One (MSXDC
G28.61-00.26) of them actually has two IRDCs overlaid in the sight of line and we
distinguish them with MSXDC G28.61-00.26(a) and MSXDC G28.61-00.26(b). And compared our
sample with the previous researches, we find that some of our IRDCs were studied in
detail at continuum emissions and molecular lines emissions
\citep[e.g.][]{sakai13,xu13,pitann13} or contained in a sample of the previous statistic
studies \citep{rathborne06,du08,sanhu12}, but the northern IRDCs have never been observed
in such large scale and number in $\rm ^{12}CO$ (1-0), $\rm ^{13}CO$ (1-0) and $\rm
C^{18}O$ (1-0) lines. This will be the first time. And we will probably to obtain some
typical characteristics and statistical properties of the IRDCs in the northern sky
through analyzing this sample.

\subsection{Survey Data}
The 1.1 mm radio continuum emission data were obtained from the
Bolocam Galactic Plane Survey
(BGPS)\footnote{http://irsa.ipac.caltech.edu/data/$\rm
BOLOCAM\_GPS$/} \citep{glenn09}. The BGPS was one of the first
large-area, systematic continuum surveys of the northern Galactic
plane in the millimeter regime, spanning the entire first quadrant
of the Galaxy with a latitude range of $\rm |b| < 0.5$ deg from the
Galactic plane and portions of the second quadrant and covering
total 220 $\rm deg^2$ at $33''$ resolution \citep{aguirre11}.

We also downloaded the 8 $\mu$m data from the Galactic Legacy
Infrared Mid-Plane Survey Extraordinaire
(GLIMPSE)\footnote{http://irsa.ipac.caltech.edu/data/SPITZER/GLIMPSE},
which was a mid-infrared survey of the inner Galaxy performed with
the Spitzer Space Telescope in a number of mid-infrared wavelength
bands at 3.6, 4.5, 5.8 and 8 $\mu$m using the Infrared Array Camera
(IRAC), which had an angular resolution between $1.5''$ and $1.9''$
\citep{fazio04,werner04}.

\section{Results}
\subsection{Morphologies of the IRDCs}
Figure 1 shows the integrated intensity maps of $\rm^{13}CO$ (1-0) and $\rm C^{18}O$
(1-0) overlaying on Spitzer 8 $\mu$m emission and on the 1.1 mm continuum emission for
each IRDC, respectively. From Figure 1, We find that the distributions of the molecule
$\rm C^{18}O$ are associated well with the IRDCs, exhibited by the high Spitzer 8 $\mu$ m
dark extinction, as well as with the 1.1 mm dust emission. Besides, 10 IRDCs are
filamentary structures and these filamentary IRDCs are fragmented into several
heterogeneous cores, but the remaining two compact IRDCs: MSXDC G31.38+00.29 and
G38.95-00.47 do not and just show one compact core. This suggests that filamentary should
be the typical shape for the IRDCs and can provide the place for the cluster-formation.
In total, 41 cores are identified according to the contour peaks in the C$^{18}$O
integrated intensity and their center positions are marked by the red crosses in the maps
and listed in Table 1. The comparison between the core centers with the peaks of the 1.1
mm dust continuum emission shows that they seem to coincide for all the cores within the
beam of our observation. This indicates that $\rm C^{18}O$ is definitely not very
optically thick. Actually it is optically thin in these IRDC cores according to the
estimation in Section 3.4 and consequently can be used to trace the dense or central
regions of the IRDC cores. While for the $\rm ^{13}CO$ (1-0) emission, they are also
associated well with the IRDCs and almost have the same shapes with the $\rm C^{18}O$
(1-0) emission but more extended. In fact, the integrated intensity maps of $\rm ^{13}CO$
do not show the cores as many as those of $\rm C^{18}O$ and the peaks of the $\rm
^{13}CO$ emission often offset a few tenths or even a few arcminutes from those of the
$\rm C^{18}O$ emission and 1.1 mm dust emission. Hence, it is probable that the molecule
$\rm ^{13}CO$ is optically thick in the IRDCs and can trace the relatively external
regions. We demonstrate this possibility in Section 3.4 and utilize it to study the
kinematics characteristics of the IRDCs.

Additionally, we examine the environment for every core and present the associated HII/UCHII regions (fill blue square) and IRAS sources (the purple stars) for them,
which are identified from SMBAD \footnote{http://simbad.u-strasbg.fr/simbad/}. You can see in Figure 1. In addition, we find a new fact that MSXDC G28.61-00.26
actually is two different IRDCs distributed in the same line of sight with the kinematical distances of 3.6 kpc and 4.6 kpc, separately. We name them to be MSXDC
G28.61-00.26(a) and MSXDC G28.61-00.26(b) and show their the integrated intensity maps in blue color for (a) and green color for (b) in Figure 1, respectively. And
MSXDC G25.04-00.20 has the same name with an IRDC in \citet{rathborne06}, but it is definitely a new source with a different morphology and distributes in a
different place.

\subsection{The extracted spectra from the centers of the IRDC cores}
The molecular spectra showed in Figure 1 are extracted from the center of each IRDC core. The $\rm ^{13}CO$ (1-0) and $\rm C^{18}O$ (1-0) lines present several
velocity components along the line of sight observations. In order to determine the systemic velocity and the velocity range of very IRDC, we draw the channel maps
in $\rm C^{18}O$ (1-0) and $\rm ^{13}CO$ (1-0) lines for each IRDC. The red dash lines in the spectra mark the systemic velocities of the IRDCs and the velocity
components between the black dash lines represent the velocity ranges associated with the IRDCs, which is listed in Table 1. From the extracted spectra, we can find
that the $\rm ^{13}CO$ (1-0) lines are much stronger than the $\rm C^{18}O$ (1-0) lines and show various shapes, such as Gauss shape, blue profile, red profile and
double components, actually indicating different dynamical information. It will be discussed in Section 4.1. We make Gauss fits to the marked velocity components
and obtain the fitting parameters of 40 IRDC cores (IRDC core MSXDC G28.61-00.26-M4 is excluded because the spectrum of its $\rm C^{18}O$ has too bad SNR to fit.
Therefore, we will not consider it in the following part.), which are listed in Table 1.

\subsection{Distance}
In section 3.2, We get the central line velocities towards the 12
northern IRDCs from the Gauss fits, respectively. Giving $\rm
C^{18}O$ is optically thin in the IRDCs, it can trace the regions
nearer to the centers of the molecular clouds than $\rm ^{13}CO$.
Therefore, we use the average $\rm V_{LSR}$ of the $\rm C^{18}O$
lines to determine the distances of the IRDCs through the rotation
curve of \citet{reid09} ($\rm R_0=8.4\, kpc$, $\rm Q_0=254\, km\,
s^{-1} $). The derived distances of the 12 northern IRDCs are
presented in Table 2 column 2 and are in the range of $(2-6)$ kpc.

\begin{table*}
\centering
\begin{minipage}{170mm}
\caption{Observed properties of the IRDCs}
 \tiny
 \begin{tabular}{lcccccccccccc}
  \hline\noalign{\smallskip}
 MSX& \multirow{2}{*}{RA}  &\multirow{2}{*}{DEC} &\multirow{2}{*}{range}& \multicolumn{4}{c}{$\rm ^{13}CO(1-0)$}  &  &\multicolumn{4}{c}{$\rm C^{18}O(1-0)$}
                \\ \cline{5-8} \cline{10-13}
ID              &                &                &           &$\rm T_{mb}$   &$\rm\int{T_{mb}dV}$ &$\rm V_{LSR}$      &$\Delta V$    &       & $\rm T_{mb}$  &$\rm\int{T_{mb}dV}$  & $\rm V_{LSR}$     &$\Delta V$    \\
(MSXDC)         & (J2000.0)      &  (J2000.0)     &$\rm(km\,
s^{-1})$& (K)      & $\rm(K \,km \,s^{-1})$& $\rm(km \,s^{-1})$
&$\rm(km\, s^{-1})$   &    & (K)     &$\rm(K\, km \,s^{-1})$
&$\rm(km \,s^{-1})$    &$\rm(km \,s^{-1})$ \\ \hline
G24.00+00.15    & 18:34:30.25    & -07:52:24.5    &   75-85        &          &               &               &                   &             &                &               &            \\
     1          & 18:34:31.67    & -07:51:31.1    &                &5.75(0.28)&   25.84(0.37) & 80.81 (0.03)  &4.29(0.07)    &     & 2.34(0.23)  &   6.61(0.30)   &  81.15 (0.04) & 2.41(0.13) \\
     2          & 18:34:24.59    & -07:53:20.9    &                &5.45(0.30)&   27.43(0.39) & 80.26 (0.03)  &4.92(0.08)    &     & 2.21(0.21)  &   8.06(0.27)   &  80.76 (0.05) & 3.32(0.13) \\
     3          & 18:34:23.70    & -07:54:35.7    &                &6.74(0.30)&   31.53(0.39) & 80.14 (0.03)  &4.84(0.09)    &     & 2.51(0.26)  &   8.45(0.34)   &  80.32 (0.06) & 3.26(0.14) \\
G25.04-00.20    & 18:37:41.26    & -07:06:39.3    &   42-50        &          &               &               &               &    &          &                &               &            \\
     1          & 18:37:28.20    & -07:09:01.1    &                &5.06(0.24)&   23.83(0.28) & 46.48 (0.03)  &4.57(0.06)     &    & 1.94(0.28)  &   7.16(0.33)   &  46.19 (0.08) & 3.60(0.15) \\
G28.61-00.26(a) & 18:44:28.25    & -03:57:52.2    &   60-65        &          &               &                &              &     &         &                &               &            \\
     1          & 18:44:25.06    & -03:57:56.6    &                &4.88(0.19)&   10.63(0.18) & 63.01 (0.01)  &2.08(0.04)     &    & 1.67(0.19)  &   2.79(0.18)   &  63.09 (0.04) & 1.53(0.09) \\
     2          & 18:44:15.23    & -04:01:53.4    &                &3.01(0.20)&   4.48(0.18)  & 63.25 (0.02)  &1.45(0.06)     &    & 0.58(0.19)  &   1.29(0.18)   &  63.22 (0.16) & 0.84(0.33) \\
     3          & 18:44:22.22    & -04:02:21.5    &                &4.23(0.25)&   6.10(0.23)  & 63.80 (0.02)  &1.33(0.04)     &    & 1.54(0.24)  &   1.85(0.22)   &  63.87 (0.05) & 0.88(0.11) \\
     4          & 18:44:24.41    & -04:01:03.0    &                &4.63(0.23)&   6.79(0.21)  & 63.97 (0.02)  &1.47(0.04)     &    & ---         &---             &---            &-----       \\
G28.61-00.26(b) & 18:44:28.25    & -03:57:52.2    &   82-95        &          &        &               &                   &  &       &                &               &            \\
     1          & 18:44:17.44    & -03:59:28.2    &                &5.70(0.24)&   29.42(0.36) & 86.40 (0.06)  &4.85(0.14)     &    & 1.83(0.23)  &   8.31(0.34)   &  86.38 (0.08) & 4.06(0.21) \\
     2          & 18:44:23.30    & -04:01:53.4    &                &5.93(0.24)&   21.96(0.36) & 88.01 (0.02)  &3.03(0.05)     &    & 2.09(0.27)  &   4.54(0.40)   &  88.03 (0.06) & 2.02(0.14) \\
     3          & 18:44:41.47    & -04:01:44.0    &                &6.92(0.39)&   26.87(0.58) & 88.53 (0.03)  &3.36(0.08)     &    & 2.54(0.25)  &   6.37(0.37)   &  88.41 (0.04) & 2.17(0.12) \\
     4          & 18:44:48.06    & -04:01:16.3    &                &5.57(0.38)&   39.43(0.56) & 89.52 (0.05)  &5.74(0.12)     &    & 2.19(0.34)  &   9.28(0.51)   &  89.29 (0.08) & 3.43(0.26) \\
     5          & 18:44:30.56    & -04:01:25.9    &                &5.71(0.27)&   19.82(0.40) & 87.46 (0.02)  &3.06(0.06)     &    & 1.71(0.18)  &   3.94(0.27)   &  87.50 (0.07) & 2.16(0.16) \\
G30.77+00.22    & 18:46:42.61    & -01:49:03.0    &   75-88        &          &        &               &                   &  &       &                &               &            \\
     1          & 18:46:47.59    & -01:48:53.2    &                &3.86(0.24)&   24.50(0.31) & 82.13 (0.05)  &6.39(0.16)     &    & 1.21(0.18)  &   4.57(0.27)   &  81.82 (0.08) & 3.63(0.21) \\
     2          & 18:46:37.01    & -01:49:08.3    &                &2.80(0.29)&   18.14(0.38) & 81.22 (0.09)  &7.23(0.24)     &    & 0.90(0.19)  &   3.41(0.28)   &  81.36 (0.12) & 3.92(0.3 ) \\
     3          & 18:46:30.54    & -01:52:05.0    &                &2.46(0.32)&   19.08(0.42) & 81.83 (0.17)  &6.53(0.17)     &    & 0.70(0.16)  &   2.54(0.24)   &  80.42 (0.12) & 2.49(0.38) \\
     4          & 18:46:26.07    & -01:53:33.7    &                &2.01(0.30)&   20.01(0.39) & 80.88 (0.14)  &4.87(0.44)     &    & 0.96(0.22)  &   3.45(0.33)   &  80.21 (0.12) & 2.35(0.25) \\
     5          & 18:46:22.03    & -01:54:05.4    &                &2.21(0.30)&   14.86(0.39) & 80.69 (0.11)  &7.93(0.32)     &    & 0.96(0.18)  &   1.59(0.27)   &  80.53 (0.14) & 2.04(0.36) \\
     6          & 18:46:44.75    & -01:53:08.0    &                &3.58(0.42)&   21.69(0.55) & 82.84 (0.17)  &6.28(0.17)     &    & 0.93(0.20)  &   5.21(0.30)   &  81.98 (0.09) & 1.73(0.26) \\
G30.97-00.14    & 18:48:21.44    & -01:48:35.6    &   73-84        &          &        &               &                   &  &       &                &               &            \\
     1          & 18:48:21.09    & -01:48:35.5    &                &5.68(0.24)&   26.16(0.33) & 77.82 (0.03)  &4.32(0.06)     &    & 1.30(0.19)  &   5.64(0.26)   &  77.72 (0.07) & 3.14(0.13) \\
     2          & 18:48:12.36    & -01:45:06.9    &                &5.09(0.37)&   40.92(0.51) & 77.96 (0.12)  &4.42(0.29)     &    & 1.31(0.28)  &   8.53(0.38)   &  77.74 (0.14) & 2.51(0.38) \\
G31.38+00.29    & 18:47:34.48    & -01:14:55.9    &   92-101.02    &          &        &               &                   &  &       &                &               &            \\
     1          & 18:47:34.39    & -01:12:52.7    &                &7.53(0.29)&   35.43(0.36) & 97.33 (0.03)  &5.45(0.09)     &    & 3.01(0.25)  &   13.55(0.31)  &  97.25 (0.05) & 4.25(0.12) \\
G31.97+00.07    & 18:49:26.07    & -00:49:30.0    &   89-102       &          &               &               &               &    &          &                &               &            \\
     1          & 18:49:35.79    & -00:46:08.3    &                &7.79(0.27)&   57.74(0.40) & 95.75 (0.03)  &7.25(0.06)     &    & 1.69(0.29)  &   10.13(0.43)  &  96.30 (0.12) & 5.88(0.24) \\
     2          & 18:49:24.45    & -00:50:49.6    &                &5.46(0.31)&   23.90(0.46) & 95.73 (0.04)  &4.12(0.09)     &    & 1.53(0.24)  &   4.59(0.36)   &  95.91 (0.1 ) & 3.05(0.22) \\
G33.69-00.01    & 18:52:52.40    & 00:40:12.4     &   97-112       &          &        &               &                   &  &       &                &               &            \\
     1          & 18:52:56.45    & 00:43:40.7     &                &4.96(0.24)&   27.90(0.38) & 106.31(0.03)  &5.30(0.08)     &    & 1.63(0.20)  &   7.50(0.32)   &  106.94(0.08) & 3.80(0.22) \\
     2          & 18:52:52.31    & 00:40:40.6     &                &4.37(0.20)&   33.55(0.32) & 105.10(0.04)  &7.36(0.08)     &    & 1.03(0.16)  &   6.68(0.26)   &  105.14(0.15) & 6.76(0.33) \\
     3          & 18:52:51.13    & 00:37:50.0     &                &6.68(0.22)&   39.99(0.35) & 104.59(0.02)  &4.92(0.06)     &    & 2.52(0.21)  &   9.04(0.34)   &  104.45(0.04) & 3.12(0.10) \\
     4          & 18:52:30.06    & 00:34:34.4     &                &3.99(0.37)&   30.91(0.59) & 103.86(0.17)  &6.44(0.17)     &    & 1.53(0.33)  &   7.81(0.53)   &  104.31(0.14) & 5.28(0.30) \\
G34.43+00.24    & 18:53:19.00    & 01:26:33.7     &   56.5-59.2      &          &        &               &                   &  &       &                &               &            \\
     1          & 18:53:14.88    & 01:31:00.0     &                &5.98(0.30)&   10.51(0.20) & 58.37 (0.03)  &2.00(0.07)     &    & 2.13(0.20)  &   2.71(0.14)   &  58.59 (0.05) & 1.32(0.13) \\
     2          & 18:53:17.19    & 01:29:30.8     &                &5.05(0.27)&   11.25(0.18) & 58.21 (0.03)  &2.15(0.07)     &    & 1.91(0.24)  &   2.88(0.16)   &  58.42 (0.05) & 1.74(0.12) \\
     3          & 18:53:18.36    & 01:27:11.9     &                &6.13(0.29)&   14.88(0.20) & 57.97 (0.04)  &3.17(0.09)     &    & 1.76(0.19)  &   3.74(0.13)   &  57.63 (0.05) & 2.81(0.11) \\
     4          & 18:53:18.64    & 01:24:13.5     &                &8.88(0.21)&   15.07(0.14) & 57.13 (0.02)  &5.07(0.06)     &    & 1.95(0.24)  &   4.39(0.16)   &  57.02 (0.10) & 4.73(0.27) \\
     5          & 18:53:19.18    & 01:22:37.4     &                &5.44(0.41)&   13.01(0.28) & 56.98 (0.07)  &3.52(0.19)     &    & 2.32(0.34)  &   4.36(0.23)   &  57.38 (0.06) & 2.10(0.16) \\
     6          & 18:53:12.86    & 01:23:28.9     &                &4.75(0.33)&   11.52(0.22) & 57.11 (0.05)  &4.25(0.15)     &    & 2.07(0.27)  &   3.12(0.18)   &  57.20 (0.16) & 2.83(0.66) \\
G38.95-00.47    & 19:04:07.50    & 05:08:18.9     &   38-45        &          &               &               &               &    &          &                &               &            \\
     1          & 19:04:07.19    & 05:09:00.3     &                &9.77(0.20)&   35.56(0.22) & 42.04 (0.01)  &3.56(0.02)     &    & 2.49(0.19)  &   6.91(0.21)   &  42.10 (0.04) & 2.66(0.08) \\
G48.52-00.47    & 19:22:08.69    & 13:36:56.6     &   35-40        &          &        &               &                   &  &       &                &               &            \\
     1          & 19:22:07.47    & 13:37:02.1     &                &3.48(0.24)&   9.68(0.22)  & 37.52 (0.03)  &2.75(0.08)     &    & 1.20(0.20)  &   2.04(0.18)   &  37.83 (0.07) & 1.77(0.14) \\
     2          & 19:22:07.02    & 13:35:17.4     &                &2.84(0.24)&   7.53(0.22)  & 37.64 (0.04)  &2.56(0.09)     &    & 0.67(0.19)  &   1.45(0.18)   &  37.67 (0.13) & 2.16(0.27) \\
G48.65-00.29    & 19:21:44.59    & 13:48:54.6     &   31-37.5      &          &        &               &                   &  &       &                &               &            \\
     1          & 19:21:34.61    & 13:51:46.9     &                &3.93(0.27)&   10.05(0.27) & 34.84 (0.03)  &2.53(0.07)     &    & 1.24(0.21)  &   2.52(0.21)   &  34.77 (0.07) & 1.85(0.18) \\
     2          & 19:21:48.31    & 13:48:50.2     &                &5.86(0.27)&   14.17(0.27) & 33.35 (0.02)  &2.25(0.05)     &    & 1.97(0.19)  &   3.23(0.19)   &  33.39 (0.04) & 1.50(0.09) \\
     3          & 19:21:46.74    & 13:45:30.7     &                &3.38(0.31)&   9.25(0.31)  & 33.62 (0.04)  &2.40(0.11)     &    & 2.42(0.11)  &   2.33(0.11)   &  34.32 (0.10) & 1.49(0.10) \\
     4          & 19:21:42.47    & 13:43:32.9     &                &3.19(0.42)&   8.92(0.42)  & 34.07 (0.06)  &2.62(0.15)     &    & 0.98(0.25)  &   2.78(0.25)   &  33.87 (0.17) & 1.95(0.37) \\

\noalign{\smallskip}\hline
       \end{tabular}
       \small{Note: Columns are (1) source name referring to \citet{Parsons09},
       (2)(3) the right ascension and declination of the reference positions and
       the core centers, (4) the velocity ranges associated with the
       IRDCs, (5)-(8) the Gauss fitting parameters of the $\rm
       ^{13}CO$ (1-0) lines, (9)-(12) the Gauss fitting parameters of the $\rm
       C^{18}O$ (1-0) lines.
        }
        \end{minipage}\\
       \end{table*}

       \begin{table*}
\centering
\begin{minipage}{170mm}
\caption{Calculated physical parameters of the IRDCs}
 \tiny
 \begin{tabular}{ccccccccccccc}
  \hline\noalign{\smallskip}

MSX(ID)       &Distance   &$\rm T_{ex}$   &R        &$\rm\tau_{_{^{13}CO}}$ &$\rm\tau_{_{C^{18}O}}$   &$\rm N({C^{18}O})$    &$\rm N({^{13}CO})$           &$\rm N(H_2)$        &$ \rm n(H_2)$          &$\rm M_{LTE}$         &$\rm M_{vir}$\\
 MSXDC               & (kpc)     &  (K)          & (pc)    &                    &                      & $\rm(10^{15}\,cm^{-2})$  & $\rm(10^{16}\,cm^{-2})$   &$\rm(10^{22}\,cm^{-2})$ &$\rm(10^{3}\,cm^{-3})$& $\rm(10^3\,M_\odot)$   & $\rm (10^3\,M_\odot)$\\\hline
G24.00+00.15    &  4.4      &               &         &               &              &              &         &             &                 &                 &              \\
     1          &           & 9.23  (0.07)  & 0.50    & 3.41          &  0.62        &  7.57 (0.36) &9.08  (0.43)              &    4.54(0.21)         &    14.8(0.69)         &    0.77(0.04)         &   0.60       \\
     2          &           & 8.90  (0.09)  & 1.11    & 3.47          &  0.63        &  9.16 (0.34) &11.00 (0.41)            &    5.50(0.20)         &    8.02 (0.30)        &    4.61(0.17)         &   2.55       \\
     3          &           & 10.22 (0.09)  & 1.47    & 3.69          &  0.67        &  10.3 (0.43) &12.34 (0.52)            &    6.17(0.26)         &    6.81 (0.29)        &    9.03(0.38)         &   3.25       \\
G25.04-00.20     &  2.9      &               &         &               &              &              &                 &                 &                        &                 &              \\
     1          &           & 8.47  (0.06)  & 1.09    & 3.58          &  0.65        &  8.09 (0.38) &9.71  (0.46)            &    4.85(0.23)         &    7.24(0.34)         &    3.88(0.18)         &   2.93       \\
G28.61-00.26(a)  &  3.6      &               &         &               &              &              &                  &                 &                 &                 &              \\
     1          &           & 8.23  (0.04)  & 0.81    & 3.91          &  0.71        &  3.21 (0.20) &3.86  (0.24)            &    1.93(0.12)         &    3.88(0.25)         &    0.85(0.05)         &   0.39       \\
     2          &           & 6.19  (0.06)  & 0.31    & 5.34          &  0.97        &  1.59 (0.22) &1.91  (0.26)            &    0.96(0.13)         &    4.94 (0.68)        &    0.06(0.009)         &   0.05       \\
     3          &           & 7.56  (0.04)  & 0.53    & 3.74          &  0.68        &  2.06 (0.25) &2.47  (0.30)            &    1.24(0.15)         &    3.76(0.45)         &    0.24(0.03)         &   0.09       \\
     4          &           &               &         &                &             &               &                       &                       &                       &                       &              \\
G28.61-00.26(b) &  4.6      &               &         &               &              &              &                  &                 &                 &                 &              \\
     1          &           & 9.08  (0.15)  & 1.59    & 4.07          &  0.74        &  9.98 (0.48) &12.00 (0.57)            &    5.99(0.29)         &    6.11(0.29)         &    10.3(0.49)         &   5.46       \\
     2          &           & 9.35  (0.05)  & 0.91    & 3.85          &  0.7         &  5.41 (0.48) &6.49  (0.58)            &    3.25(0.29)         &    5.80(0.52)         &    1.82(0.16)         &   0.77       \\
     3          &           & 10.40 (0.08)  & 0.98    & 3.74          &  0.68        &  7.84 (0.47) &9.41  (0.56)            &    4.70(0.28)         &    7.82(0.47)         &    3.03(0.18)         &   0.96       \\
     4          &           & 9.02  (0.13)  & 1.08    & 3.52          &  0.64        &  10.6 (0.62) &12.80 (0.75)            &    6.38(0.37)         &    9.57(0.56)         &    5.07(0.30)         &   2.65       \\
     5          &           & 9.07  (0.06)  & 0.70    & 4.24          &  0.77        &  4.79 (0.33) &5.75  (0.40)            &    2.88(0.20)         &    6.72(0.46)         &    0.94(0.06)         &   0.68       \\
G30.77+00.22    &  4.3      &               &         &               &              &              &                  &                 &                 &                 &             \\
     1          &           & 7.13  (0.17)  & 0.96    & 4.13          &  0.75        &  5.20 (0.37) &6.24  (0.44)            &    3.12(0.22)         &    5.26(0.37)         &    1.96(0.14)         &   2.64       \\
     2          &           & 6.00  (0.26)  & 0.55    & 4.07          &  0.74        &  3.81 (0.43) &4.57  (0.51)            &    2.29(0.26)         &    6.75(0.76)         &    0.47(0.05)         &   1.76       \\
     3          &           & 5.62  (0.19)  & 0.64    & 4.40          &  0.8         &  2.92 (0.32) &3.51  (0.39)            &    1.75(0.19)         &    4.46(0.50)         &    0.48(0.05)         &   0.82       \\
     4          &           & 5.21  (0.50)  & 0.58    & 2.97          &  0.54        &  3.57 (0.73) &4.28  (0.87)            &    2.14(0.44)         &    6.04(1.23)         &    0.48(0.10)         &   0.66       \\
     5          &           & 5.41  (0.36)  & 0.40    & 3.25          &  0.59        &  1.67 (0.35) &2.01  (0.42)            &    1.00(0.21)         &    4.07(0.85)         &    0.11(0.02)         &   0.35       \\
     6          &           & 6.81  (0.18)  & 0.51    & 4.62          &  0.84        &  6.12 (0.44) &7.35  (0.53)            &    3.67(0.27)         &    11.6(0.84)         &    0.65(0.05)         &   0.32       \\
G30.97-00.14    &  4.2      &               &         &               &              &              &                  &                 &                 &                 &              \\
     1          &           & 9.00  (0.06)  & 0.83    & 4.95          &  0.9         &  7.24 (0.34) &8.69  (0.41)            &    4.34(0.21)         &    8.49(0.40)         &    2.03(0.10)         &   1.70       \\
     2          &           & 8.40  (0.30)  & 1.01    & 4.62          &  0.84        &  10.5 (0.74) &12.50 (0.89)            &    6.27(0.45)         &    10.0 (0.72)        &    4.36(0.31)         &   1.33       \\
G31.38+00.29    &  5.1      &               &         &               &              &              &                  &                 &                 &                 &              \\
     1          &           & 11.10 (0.09)  & 1.28    & 3.47          &  0.63        &  1.68 (0.43) &20.10 (0.52)            &    10.1(0.26)         &    12.8(0.33)         &    11.1(0.29)         &   4.80       \\
G31.97+00.07    &  5.1      &               &         &               &              &              &                  &                 &                 &                 &              \\
     1          &           & 11.17 (0.06)  & 1.68    & 5.06          &  0.92        &  14.3 (0.62) &17.10 (0.74)            &    8.56(0.37)         &    8.30(0.36)         &    16.3(0.70)         &   12.1       \\
     2          &           & 8.80  (0.09)  & 0.62    & 4.40          &  0.8         &  5.60 (0.45) &6.72  (0.53)            &    3.36(0.27)         &    8.76(0.70)         &    0.88(0.07)         &   1.21       \\
G33.69-00.01    &  5.8      &               &         &               &              &              &                  &                 &                &                 &              \\
     1          &           & 8.31  (0.08)  & 0.99    & 4.02          &  0.73        &  8.73 (0.40) &10.50 (0.48)            &    5.24(0.24)         &    8.55(0.39)         &    3.51(0.16)         &   2.99       \\
     2          &           & 7.64  (0.08)  & 1.25    & 4.84          &  0.88        &  8.14 (0.34) &9.76  (0.41)            &    4.88(0.21)         &    6.35(0.27)         &    5.14(0.22)         &   11.9       \\
     3          &           & 10.17 (0.06)  & 1.32    & 3.63          &  0.66        &  10.9 (0.42) &13.10 (0.50)            &    6.56(0.25)         &    8.09(0.31)         &    7.67(0.29)         &   2.67       \\
     4          &           & 7.32  (0.18)  & 1.03    & 3.63          &  0.66        &  8.57 (0.67) &10.30 (0.81)            &    5.14(0.40)         &    8.12(0.61)         &    3.68(0.29)         &   5.97       \\
G34.43+00.24    &  3.3      &               &         &               &              &              &                  &                 &                 &                 &             \\
     1          &           & 9.41  (0.07)  & 0.57    & 3.80          &  0.69        &  3.22 (0.17) &3.87  (0.20)            &    1.93(0.10)         &    5.55(0.29)         &    0.42(0.02)         &   0.21       \\
     2          &           & 8.45  (0.07)  & 0.27    & 3.63          &  0.66        &  3.27 (0.19) &3.92  (0.23)            &    1.96(0.11)         &    11.8(0.69)         &    0.10(0.006)        &   0.17       \\
     3          &           & 9.50  (0.09)  & 0.42    & 4.35          &  0.79        &  4.66 (0.17) &5.60  (0.21)            &    2.80(0.11)         &    10.8(0.40)         &    0.34(0.01)         &   0.69       \\
     4          &           & 12.29 (0.06)  & 0.97    & 5.01          &  0.91        &  6.45 (0.24) &7.74  (0.29)            &    3.87(0.15)         &    6.48(0.24)         &    2.46(0.09)         &   4.51       \\
     5          &           & 9.07  (0.45)  & 0.51    & 2.81  　　　　&　0.51        &  4.76 (0.29) &5.71  (0.52)            &    2.85(0.26)         &    9.11(0.83)         &    0.50(0.05)         &   0.47       \\
     6          &           & 8.28  (0.36)  & 0.68    & 2.92 　　　　 &　0.53　　　　&  4.64 (0.28) &5.57  (0.44)            &    2.78(0.22)         &    6.63(0.53)         &    0.87(0.07)         &   1.14       \\
G38.95-00.47    &  2.5      &               &         &               &              &              &                 &                 &                 &                 &              \\
     1          &           & 13.22 (0.02)  & 0.70    & 4.68          &  0.85        &  10.3 (0.31) &12.34 (0.37)            &    6.17(0.19)         &    14.3(0.43)         &    2.03(0.06)         &   1.03       \\
G48.52-00.47    &  2.5      &               &         &               &              &              &                 &                 &                 &                 &             \\
     1          &           & 6.74  (0.09)  & 0.44    & 3.91          &  0.71        &  2.26 (0.21) &2.72  (0.25)            &    1.36(0.13)         &    5.06(0.47)         &    0.18(0.02)         &   0.28       \\
     2          &           & 6.01  (0.10)  & 0.25    & 4.84          &  0.88        &  1.72 (0.21) &2.07  (0.26)            &    1.03(0.13)         &    6.59(0.82)         &    0.04(0.006)         &   0.25       \\
G48.65-00.29    &  2.2      &               &         &               &              &              &                 &                 &                 &                 &              \\
     1          &           & 7.21  (0.07)  & 0.41    & 4.13          &  0.75        &  2.87 (0.25) &3.45  (0.30)            &    1.72(0.15)         &    6.83(0.59)         &    0.20(0.02)         &   0.29       \\
     2          &           & 9.26  (0.05)  & 0.51    & 3.96          &  0.72        &  3.87 (0.23) &4.65  (0.28)            &    2.32(0.14)         &    7.37(0.44)         &    0.41(0.02)         &   0.24       \\
     3          &           & 7.53  (0.15)  & 0.36    & 1.54          &  0.28        &  2.16 (0.12) &2.59  (0.15)            &    1.29(0.07)         &    5.76(0.33)         &    0.12(0.007)         &   0.17       \\
     4          &           & 6.42  (0.16)  & 0.46    & 4.18          &  0.76        &  3.14 (0.32) &3.77  (0.38)            &    1.88(0.19)         &    6.64(0.67)         &    0.27(0.03)         &   0.37       \\

\noalign{\smallskip}\hline
       \end{tabular}
       \small{Note: The numbers in () represent the errors of the
       corresponding physical parameters.
        }
        \end{minipage}\\
       \end{table*}

\subsection{Properties of the dense IRDC cores}
In order to study the properties of the molecular cores, we calculate the physical
parameters (the excited temperature $\rm T_{ex} $, optical depth $\tau$, column density,
radius, number density and mass, etc.) of various cores. In calculations, assume that the
cores are in the state of local thermodynamical equilibrium (LTE). The molecular cores
are generally optically thin for the $\rm^{13}CO$ and $\rm C^{18}O$ emission, but
referring to the high column density $\rm\sim 10^{23}\,cm^{-2}$ and number density
$\rm\geq10^5\,cm^{-3}$ within the IRDC cores in the previous researches
\citep{egan98,carey98,carey00, rathborne06}, here we can assume that the $\rm^{13}CO$
(1-0) transition line is optically thick and the $\rm C^{18}O$ (1-0) line is optically
thin. Referring to \citet{Myers83}, we can derived the optical depths of the $\rm
^{13}CO$ and $\rm C^{18}O$ lines from following equation:
\begin{equation}
\rm \frac{T_{mb}(^{13}CO)}{T_{mb}(C^{18}O)}\simeq \frac{1-exp(-\tau_{_{^{13}CO}})}{1-exp(-\tau_{_{C^{18}O}})},
\frac{\tau_{_{^{13}CO}}}{\tau_{_{C^{18}O}}}=\frac{[^{13}CO]}{[C^{18}O]}\approx5.5
\end{equation}

The calculated optical depths of the $\rm ^{13}CO$ (1-0) and $\rm
C^{18}O$ (1-0) lines in the peak positions of the IRDC cores are
listed in Table 2 column 5, 6. We can see that the $\rm ^{13}CO$
(1-0) line is really optically thick in all the IRDC cores and the
hypothesis is valid. So the excitation temperature $\rm T_{ex}$ of
the cores can be worked out by equation (2).
\begin{equation}
\rm T_{ex}=\frac{5.29}{ln[1+5.29/(T_{mb}(^{13}CO)+0.89)]}
\end{equation}

And the column density of $\rm C^{18}O$ molecule can be derived from
the following formulae \citep{scoville86}:
\begin{equation}
\rm
\frac{N(C^{18}O)}{cm^{-2}}=4.77\times10^{13}\frac{T_{ex}+0.88}{exp(-5.27/T_{ex})}\frac{\tau}{1-exp(-\tau)}\frac{\int{T_{mb}}d\upsilon}{K\,
km \,s^{-1}}
\end{equation}

As for the below analysis, we assume that the cores are spherical and identified within
the 70$\%$ contours of the integrated intensity distribution. The characteristic size R
mean the radii of the cores which are defined as $\rm R=\sum{r_i}/\sum{i}$, where $\rm
r_i$ is the size of the core in different directions. The final results are listed in
Table 2 column 4.

Taking the element abundance ratios $\rm
N(H_2)/N(^{13}CO)=5\times10^5$ and $\rm
N(H_2)/N(C^{18}O)=6\times10^6$ \citep{frerking82}, and combining the
above characteristic sizes of the cores, the derived mean volume
densities of $\rm H_2$ molecule are

\begin{equation}
\rm n(H_2)=N(H_2)/2R
\end{equation}

Based on the LTE assumption and spherical model, the masses of the cores are calculated as:
\begin{equation}
\rm M_{LTE}=\mu m_{H_2}n(H_2)\times(\frac{4}{3}\pi R^3)
\end{equation}
where, $\rm m_{H_2}$ is the mass of hydrogen molecule, $\mu=1.36$ is the mean molecular
weight considering the contributions of He and other heavy elements to the total mass.

In addition, the viral masses of the IRDC cores are also calculated
by using equation (6).
\begin{equation}
\rm M_{vir}=5(\Delta \upsilon)^2R/(8ln2\times G)
\end{equation}
in which $\Delta \upsilon$ is the full width at the half-maximum, G
is the gravitational constant.

All the calculated physical parameters are listed in Table 2. We plot the histograms of
all of them in Figure 2 to study the physical properties in the northern IRDCs. From
Figure 2, we can find that the northern IRDCs have a typical excitation temperature
$(8\sim10)$ K with a mean $8.4\pm1.8$ K, a maximum 13.2 K and a minimum 5.2 K, which are
lower than the previous results derived from other denser molecules (eg, $\rm NH_3$, $\rm
HCO^+$, HCN, etc.) \citep{ragan06,pillai06,liu13}, but consistent with the excitation
temperatures derived from CO \citep{wu05,du08}. This suggests that different molecules
maybe trace different cloud regions and CO seems to prefer to the colder areas and/or the
molecular emission in our observations are beam diluted. The typical value of the
integrated intensity ratio of $\rm ^{13}CO$ to $\rm C^{18}O$ being in the range $3\sim6$
with a smaller fluctuation agrees well with \citet{meier01}. Its average value is
$4.4\pm1.3$. The mean line widths of the molecules $\rm ^{13}CO$ and $\rm C^{18}O$ are
$\rm(4.3\pm1.8)\,km\,s^{-1}$ and $\rm(2.8\pm1.3)\,km\,s^{-1}$, respectively, which are
much greater than their mean thermal broadening: $\rm\Delta
V_{therm}=\sqrt{kT_{ex}/m_H\mu}\approx0.16\,km\,s^{-1}$. This indicates that the
non-therm motions will be charged for the velocity dispersion and we consider the
turbulence as the dominant.

\citet{vasy09} suggested the typical column density ranges between 0.9 and
$4.6\times10^{22}\, \rm cm^{-2}$ and mass range $\rm (50-1000)\, M_{\odot}$ in the
southern IRDCs, which are very consistent with our results for the northern IRDCs,
implying that some properties of the IRDCs in southern and in northern should be similar
on the whole. The median values of $\rm N_{H_2}$ and $\rm M_{LTE}$ are
$3.2\times10^{22}\,\rm cm^{-2}$ and $\rm 875\, M_{\odot}$. Besides, from the histograms
of $\rm\tau_{_{C^{18}O}}$, $\rm \tau_{_{^{13}CO}}$ and R in Figure 2, we can see that the
typical optical depth of $\rm C^{18}O$ is in the range of $0.6\sim0.8$ with a mean value
$0.7\pm0.1$, while for $\rm ^{13}CO$, it is in the range of $3.5\sim4.5$ with a mean
value $4.3\pm1.8$ and the typical characteristics size of the IRDC cores is from 0.5 pc
to 1.2 pc with a mean value $0.8\pm0.4$ pc. The histogram of the volume density $\rm
n(H_2)$ shows a typical range of $(6-10)\times10^3\,\rm cm^{-3}$, confirming the
viewpoint that the IRDC cores are dense. At the same time, we find that the IRDC cores
associated with the HII/UCHII regions or IRAS sources have higher excitation temperatures
more than 10 K and are more compact. This indicates that these IRDC cores are evolving
into the different stages and even some have been to the later stages and may be heated
by the associated HII/UCHII regions.

Comparing $\rm M_{LTE}$ with $\rm M_{vir}$ for the IRDC cores in Table 2, we find that
$57.5\%$ of the IRDC cores have $\rm  M_{LTE}> M_{vir}$, and consequently we suggest that
these IRDC cores are likely gravitational bounded and might be in the state of
gravitational collapse, providing a probability to form stars. In addition, we also make
a graph between $\rm M_{LTE}$ and $\rm M_{vir}$ for the 40 cores, which is showed in
Figure 3. Here, we just consider the error of $\rm M_{LTE}$ for each IRDC core, without
considering that of $\rm M_{vir}$. Because the errors caused by the line widths are too
small and thus can be ignored for $\rm M_{vir}$. Certainly, there can be significant
systematic uncertainties for the estimation of $\rm M_{vir}$, which are difficult to be
constrained and hence are not considered here. The red line in Figure 3 represents the
relation of the equality. Obviously, from this figure, we find that a large portion of
the IRDC cores seems have $\rm M_{LTE}\approx M_{vir}$, especially for the relative
low-mass source. In fact, except for the $57.5\%$ cors, up to $30\%$ of the remaining
cores have the $\rm M_{LTE}$ slightly less than the viral equilibrium mass, which we
think they are in a "metastable state". Once they suffer from the external pressure, such
as the stellar wind from the young cluster around the core, they will star the
gravitational collapse and have a further chance to form stars. Moreover, Seeing Table 2
and Figure 3, it is obvious that the uncertainties in the mass estimates are small and
thus almost have no impact to the above analysis on the stability of the cores. Besides,
comparing the gravitational cores with the other in Figure 1 and Table 2, we can find
that the $57.5\%$ cores are more compact, warmer and higher column density, implying that
they may be in the relatively later evolutional stages of the molecular cores.

\section{Discussion}
\subsection{infall and outflow}
Seeing the spectra in Figure 1, we find that the optically thick lines $\rm ^{13}CO$
(1-0) show the self-absorbed features or the "blue profiles"
\citep{Anglada87,Adelson88,Zhou92,Walker94} in three IRDC cores: MSXDC G24.00+00.15-3,
MSXDC G31.38+00.29-1 and MSXDC G34.43+00.24-4 (Later we will simplify them for G24.00-3,
G31.38-1 and G34.43-4). While their optically thin lines $\rm C^{18}O$ (1-0) just show a
single velocity component peaked at the $\rm^{13}CO$ (1-0) lines center. This satisfies
the classical signature of infall, rotation or outflows \citep{Adelson88}. However,
rotation and outflows should produce approximately equal numbers of red and blue
asymmetric profiles. On the other hand, infall ought to preferentially produce blue
asymmetric profiles \citep[e.g.][]{ Anglada87,Zhou92,Walker94,Jimenez14}. In order to
identify which one is the main culprit for that, we plot the map grids of them using the
two lines $\rm ^{13}CO$ (1-0) (Green) and $\rm C^{18}O$ (1-0) (Black), presented in
Figure 4. Their map grids indeed exhibit the large-scaled (at least $2'\times2'$) blue
asymmetric feature, which suggests that they are real infall candidates
\citep{Wu05,Ren12}.

Alternatively, \citet{Mardones97} put forward a quantization parameter $\rm \delta V$ of
the line asymmetry to identify the infall motion, which was widely used in the
astronomical researches \citep[e.g.][]{Fuller05,Wu05,Chen13}. He defined $\rm \delta V$
as $\rm \delta V=(V_{thick}-V_{thin})/\Delta V_{thin}$, where $\rm V_{thick}$ represents
the line peak velocity of the optically thick line and $\rm V_{thin}$, an optically thin
tracer, measure the systemic velocity. $\rm\Delta V_{thin}$ is the line width of the
optically thin line. \citet{Mardones97} suggested a criterion $\rm|\delta V|>0.25$ to
indicate that a line profile was asymmetric and $\rm\delta V>0.25$ for red asymmetry or
$\rm\delta V<-0.25$ for blue asymmetry. Here, we also adopt this quantitative method as
\citet{Fuller05} to further confirm our results. The calculated $\rm \delta V$ for
G24.00-3, G31.38-1 and G34.43-4 are $-0.44\pm0.02$, $-0.29\pm0.03$, and $-0.25\pm0.03$,
respectively. The uncertainty estimates of $\rm \delta V$ are caused by the velocity
dispersion of the optically thin line, i.e., $\rm\sigma_{_{V_{thin}}}$. This indicates
that there are indeed infall motions in these three cores.

Furthermore, we make Position-Velocity diagrams to identify their outflows,
finding that only the PV diagram of IRDC core G34.43-4 (in Figure 4) shows the velocity
gradients in the velocity ranges of $52\sim56.5$ km s$^{-1}$ and $59.2\sim 62.2$ km
s$^{-1}$, which implies the existence of the bipolar outflows. We present the integrated
intensity maps of the outflows with the $^{13}$CO J=1-0 line in the bottom of Figure 4.
The velocity component of blueshifted is from 52 km s$^{-1}$ to 56.5 km s$^{-1}$, while
the velocity component of redshifted is from 59.2 km s$^{-1}$ to 62.2 km s$^{-1}$. The
blueshifted and redshifted components are shown as blue and red contours. From Figure
4(Bottom), we find a big difference between the red outflow and the blue outflow. The red
outflow shows a high collimation from north to south, but its blue outflow has a badly
collimation. Their big difference in collimation probably implies that they come from
different outflows and thus have different driving sources. That we do not detect their
corresponding outflows is probable due to the low resolution of our telescope or
contamination. Besides, we also find that both of the outflows are associated well with
the molecular core G34.43-4, an UCHII region GAL 034.4+00.23 with a local standard rest
velocity 57.3 km s$^{-1}$(the blue box in Figure 4) and IRAS 18507+0121(the green star in
Figure 4) \citep{bronfman96}. These suggest that there likely be several stars or a
stellar cluster forming in IRDC core G34.43-4. Thus, we expect the high resolution
observations to make clear the internal structure of this core.

For these three IRDC cores, they were well studied by previous
researchers. IRDC core G24.00-3 was associated with IRAS 18317-0757
and an UCHII region \citep{wood89}. The high resolution observation
of \citet{hunter04} suggested considerable fragmentation having
taken place in the molecular cloud and star formation maybe ongoing
throughout the core. But nobody has detected the infall and outflow
motions before. For IRDC core G31.38-1, it was studied to be
associated with IRAS 18449-0115 and an UCHII region
\citep{cesaroni98}. \citet{gaume87} and \citet{cesaroni11} found a
striking velocity gradient across the core in the NE-SW direction
and \citet{klaassen12} suggested the infall motions in the molecular
cloud but didn't find the outflows. IRDC core G34.43-4, which
corresponds to the southern compact molecular core of
\citet{shepherd04}, was associated with IRAS 18507+0121, with an NIR
cluster of young stars with a central B0.5 star \citep{shepherd04},
with an UCHII region \citep{Miralles94,Molinari98}, with a variable
$\rm H_2O$ maser \citep{Miralles94}, with $\rm CH_3OH$ maser
emission \citep{szymczak00}, and with three massive outflows
\citep{shepherd07}. \citet{sanhueza10} suggested that it was
undergoing collapsing and had outflows.

However, it is the first time that the infall motions have been
detected in IRDC core G24.00-3 and IRDC core G34.43-4. And the
numbers and morphologies of the outflows of IRDC core G34.43-4 in
our work are different from that of \citet{shepherd07} and
\citet{sanhueza10}, which is possible to be related to the different
integrated velocity ranges for the wings and a larger spatial
resolution for our observations. But we have clearly observed a high
collimation red outflow and a compact blue outflow in $\rm^{13}CO$
(1-0) that have never been studied before. They probably also
suggest two different outflows and imply several stars forming in
IRDC core G24.00-3.

\subsection{Core mass functions}
From Table 2, we obtain a mass range of $\rm 40\,M_{\odot}~1.7\times10^4 \,M_{\odot}$, which we use to estimate the mass spectrum of the IRDC cores. The mass
spectrum is calculated from the number of the cores $\rm \Delta N$ per each mass bin $\rm \Delta M$, that is:
\begin{equation}
 \rm f(M)=\frac{d\,N}{d\,M}=\frac{\Delta N}{\Delta M}
 \end{equation}
The binned mass spectrum of the 40 IRDC cores is showed in the left
plane of Figure 5, which obeys to a power law function, with a
power-law index of $0.79\pm0.03$. Here, we do not take into account
the errors in the mass estimation but the simply counting errors:
$\rm\sigma=\sqrt{\Delta N}/\Delta M$. Moreover, we also perform a
mass spectrum to the gravitational bound cores, which is presented
in the right plane of Figure 5. It is also a power-law spectrum with
a slope of $0.83\pm0.02$, which is almost the same with that of the
whole cores, implying the possibility that the evolution of the IRDC
cores might have no influence on their mass spectrum and hence the
mass spectrum of the IRDC cores can be used to estimate the initial
mass function. However, we compare them with the power-law index of
the stellar initial mass functions (IMFs) derived from
\citet{salpeter55,scalo86,Kroupa01,Kroupa02,Chabrier03} as well as
the CO cloud mass distribution $\alpha=1.6$ \citep{solomon87},
finding that our results are much more flatten, which maybe suggest
that the IRDC cores account for a large proportion of the massive
molecular cores and therefore the IRDCs are probable to be the
birthplace of the high-mass stars/clusters. But we have reservations
about this conclusion. The large difference between our indexes with
those of others possibly comes from our small sample and the
telescope, which can cause a big error \citep{Rosolowsky05}. As for
the error coming from mass estimation, we think that it is equal to
all the researchers and it can not bring a great discrepancy for us.
Thereby, a big sample and high resolution observations are needed to
continue the study of the mass spectrum of the IRDC cores.

\subsection{Comparison with the Planck cores}
Planck satellite working at submillimeter/millimeter bands have
detected 10,783 cold clumps in its survey \citep{Planck
Collaboration11a} and therefore provided a wealth of early sources
to be studied. Actually, \citet{Planck Collaboration11a} suggested
that the planck cold clumps in the clod core Catalogue of Planck
Objects (C3PO) had dust temperatures of $(10-15)$ K and column
densities of $\sim (0.1-1.6)\times10^{22} \,\rm cm^{-2}$ and the
research of \citet{wu12} showed that the planck clumps in the Early
Cold clump Catalogue (ECC) had excitation temperatures range from 4
to 27 K and column densities $\rm N_{H_2}$ in the range of
$1.0\times10^{22}-4.5\times10^{22}\,\rm cm^{-2}$ with an average
value of $(4.4\pm3.6)\times10^{21}\,\rm cm^{-2}$. In addition, the
mean excitation temperatures and the mean column densities of the
planck cold clumps in Orion complex are in the range of $(7.4-21.1)$
K with an average $12.1\pm3.0$ K and $(0.5-9.5)\times10^{21}\,\rm
cm^{-2}$ with an average $\rm (2.9\pm1.9)\times10^{21}\, cm^{-2}$.
Their dense cores had an mean radius and LTE mass of $0.34\pm0.14$
pc and $38^{+5}_{-30}\rm \,M_{\odot}$, respectively \citep{liu12}.
Comparing them with our results, we can see that the northern IRDC
cores are colder but denser, and have larger size and much more
massive, suggesting that the IRDC cores appear to be more suitable
for the research to the early stage of massive star formation.

Moreover, \citet{liu12} did a core mass function toward 82 cores in
51 planck cold clumps projected on Orion complex and derived a
power-law index of $1.32\pm0.08$, which is larger than
$0.79\pm0.03$. Given the fact that we use the same telescope and the
same molecular lines, the difference will be caused by several
factors: 1) the size of the sample, 2)the completeness limit, 3)the
IRDC cores might have a flatten CMF than the planck cores. However,
just consider the mass distribution on the whole, the IRDC cores
indeed prefer to fragment into massive cores and provide ideal
objects for the research of the high-mass star formation, but the
planck cores are more perfect to study the early stages of the low
mass star formation.

\section{Summary}
We perform a mapping observation in molecular lines of $\rm ^{13}CO$
and $\rm C^{18}O$ $\rm J=1-0$ towards 12 northern IRDCs. The BGPS
1.1 mm data and GLIMPSE Spitzer IRAC $8\,\mu m$ data are also used.
Their morphologies and properties are analyzed. The star formation
motions and core mass function are discussed. The main findings in
this work are as follows.

1. The $\rm C^{18}O$ molecule emission, BGPS 1.1 mm emission and
Spitzer $8\,\mu m$ emission are associated well with each other and
show the similar morphologies for all the 12 northern IRDCs. And ten
IRDCs are filamentary structure containing at least one core and the
remaining IRDCs are compact and isolated. 41 dense cores are
identified in the IRDCs.

2. The typical excitation temperature $\rm T_{ex}$ is in the range of $8\sim10$ K, with an average value $8.4\pm1.8$, which is lower than the previous results
studied by denser molecules, implying that different molecules trace different regions. The typical $\rm H_2$ column density and LTE mass are $(1\sim6)\times10^{22}
\,\rm cm^{-2}$ and $60-2000 \,\rm M_{\odot}$, respectively, consistent with the typical values of southern IRDCs. Their median values are $3.2\times10^{22}\,\rm
cm^{-2}$ and $\rm 875 \,M_{\odot}$, respectively. The typical values of $\rm I_{^{13}CO}/I_{C^{18}O}$, $\rm \tau_{_{C^{18}O}}$, $\rm n(H_2)$ and R are $3\sim6$,
$0.6\sim0.8$, $(6-10)\times10^3\,\rm cm^{-3}$ and $(0.5\sim1.2)$ pc, with the corresponding average values $4.4\pm1.3$, $0.8\pm0.1$, $(8.1\pm3.2)\times10^3\,\rm
cm^{-3}$ and $0.8\pm0.4$, respectively.

3. Through the comparison with the studies of the planck cores, we
find that the IRDC cores are colder, denser, more massive and with
larger size, suggesting that the IRDC cores are more suitable for
exploring the massive star formation.

4. Three IRDC cores G24.00-3, G31.38-1 and G34.43-4 are found to
have large scaled infall motions. And IRDC core G34.43-4 probably
have two different outflows, one of which is high collimation.

5. The core mass function can be fitted with a power-law for $40<\rm
M_{LTE}/M_{\odot}<17000$, whose slope is $0.79\pm0.03$. While the
CMF of the the gravitational bound cores almost have the same
distribution function, indicating that the evolution of the IRDC
cores probably have no compact on the core mass function. Hence, it
will be a effective method to study the stellar initial mass
function. Certainly, a large sample of the IRDC cores is need to
prove this conclusion.

\section*{Acknowledgments}
%We thank the anonymous referee for whose constructive suggestions.
We are grateful to the staff at the Qinghai Station of PMO for their assistance during the observations. Thanks for the Key Laboratory for Radio Astronomy, CAS to
partly support the telescope operating. This work also has made use of the data products from the Bolocam Galactic Plane Survey (BGPS) and NASA/IPAC Infrared
Science Archive, which is operated by the Jet Propulsion Laboratory, California Institute of Technology, under contract with the National Aeronautics and Space
Administration. This work is supported by the National Natural Science Foundation of China (Grant No. 11363004).

\clearpage
\begin{figure*}
\vspace{15mm}
\centering
\includegraphics[angle=0,width=5.5in]{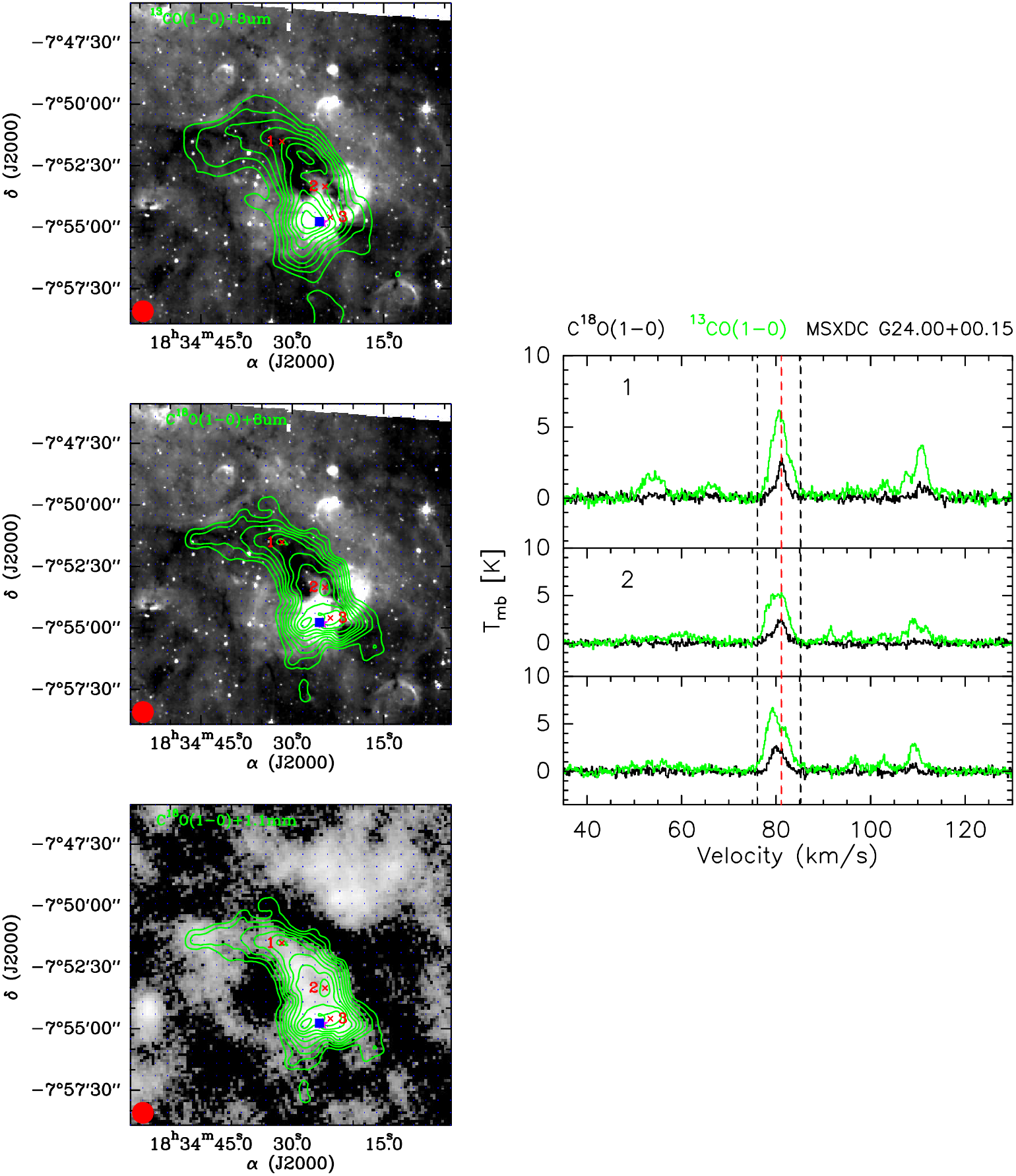}
\vspace{20mm}\caption{Top left: the $\rm ^{13}CO$ (1-0) integrated
intensity map overlays on Spizter $8 \mu m $ emission. Middle left:
the $\rm C^{18}O$ (1-0) integrated intensity map overlays on Spizter
$\rm 8 \mu m$ emission. Bottom left: the $\rm C^{18}O$ (1-0)
integrated intensity map overlays on the $\rm 1.1mm$ continuum line
emission. The blue boxes represents the HII regions, the purple stars
indicate the positions of IRAS sources. The "$\times$" marks the
centers of the labeled cores. The beam size is showed in the bottom
left corner of each diagram. Right: the $\rm ^{13}CO$ (1-0) (green)
and $\rm C^{18}O$ (1-0) (black) lines are extracted from the
"$\times$" positions of the labeled cores. The two black dash lines
show the integrated velocity ranges of the integrated intensity and
the red dash lines mark the systemic velocities towards 12 northern
IRDCs, respectively. Note MSXDC G28.61-00.26, it contains two different
IRDCs. The blue contours are for MSXDC G28.61-00.26(a) and the green contours
represent MSXDC G28.61-00.26(b).}
\end{figure*}

\setcounter{figure}{0}
 \clearpage
\begin{figure*}
\vspace{10mm} \centering
\includegraphics[angle=0]{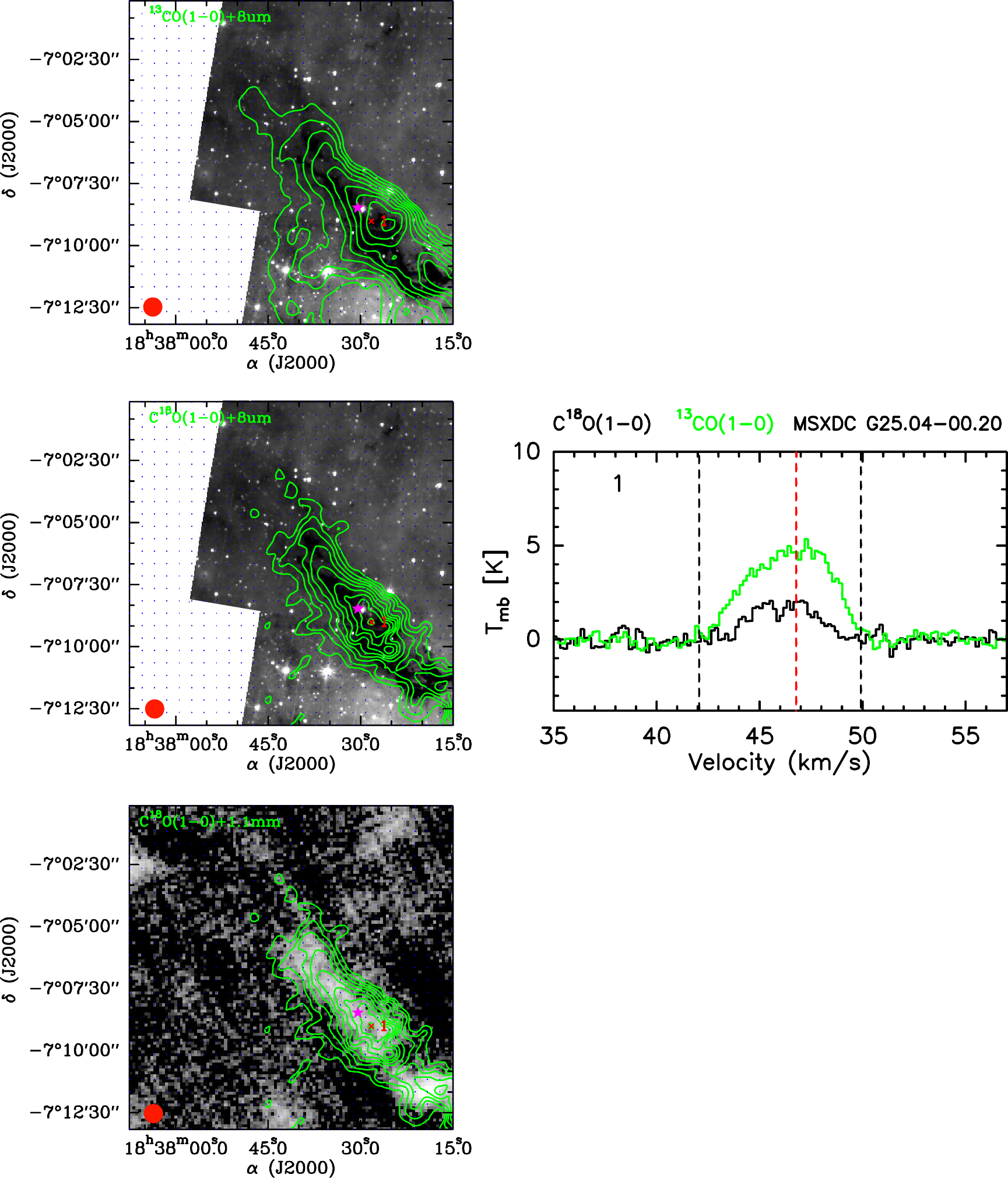}
\vspace{10mm}\caption{Continued}
\end{figure*}

\setcounter{figure}{0} \clearpage
\begin{figure*}
\vspace{15mm}
 \centering
\includegraphics[angle=0]{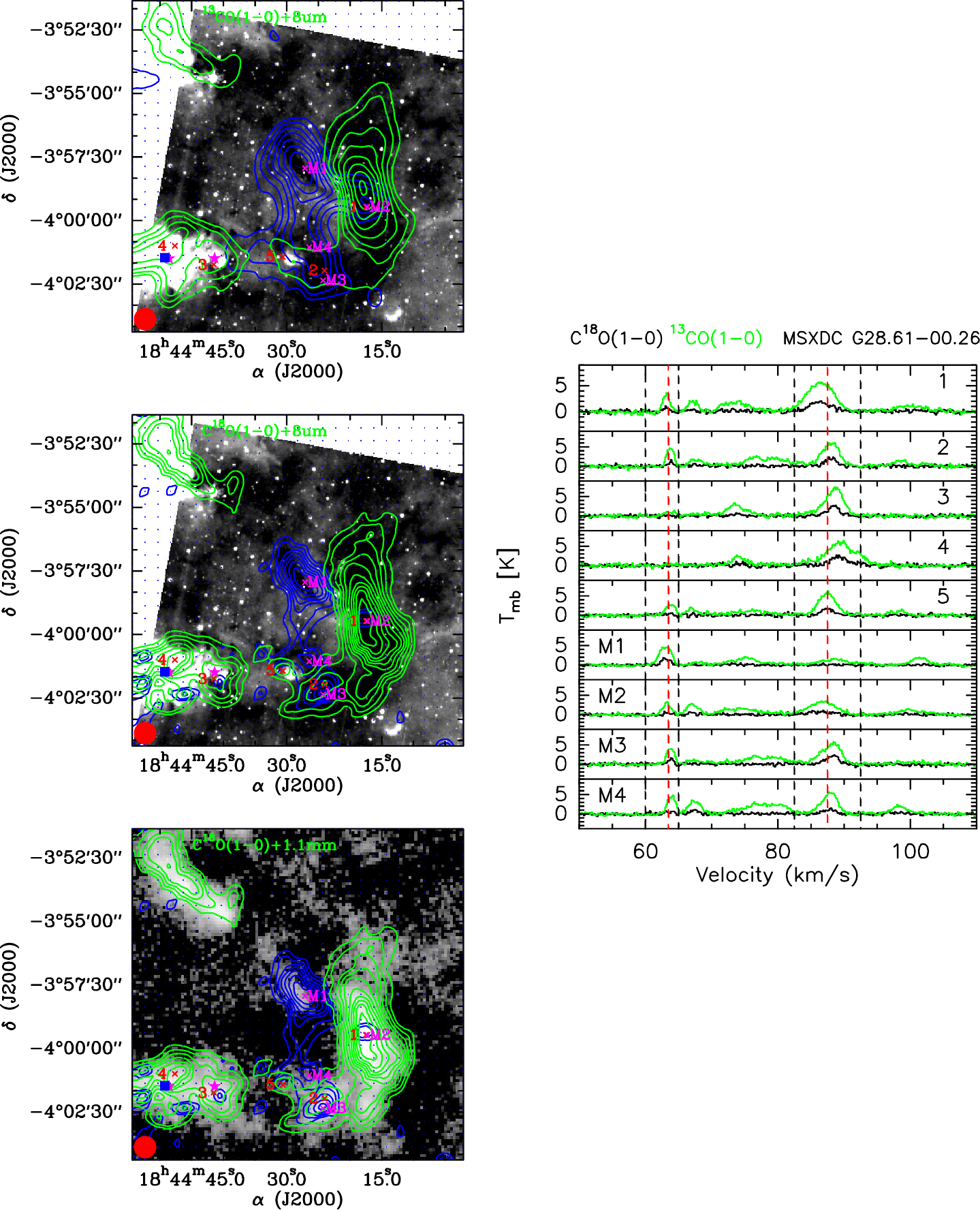}
\vspace{10mm}\caption{Continued}
\end{figure*}

\setcounter{figure}{0}
 \clearpage
\begin{figure*}
\vspace{10mm}
 \centering
\includegraphics[angle=0]{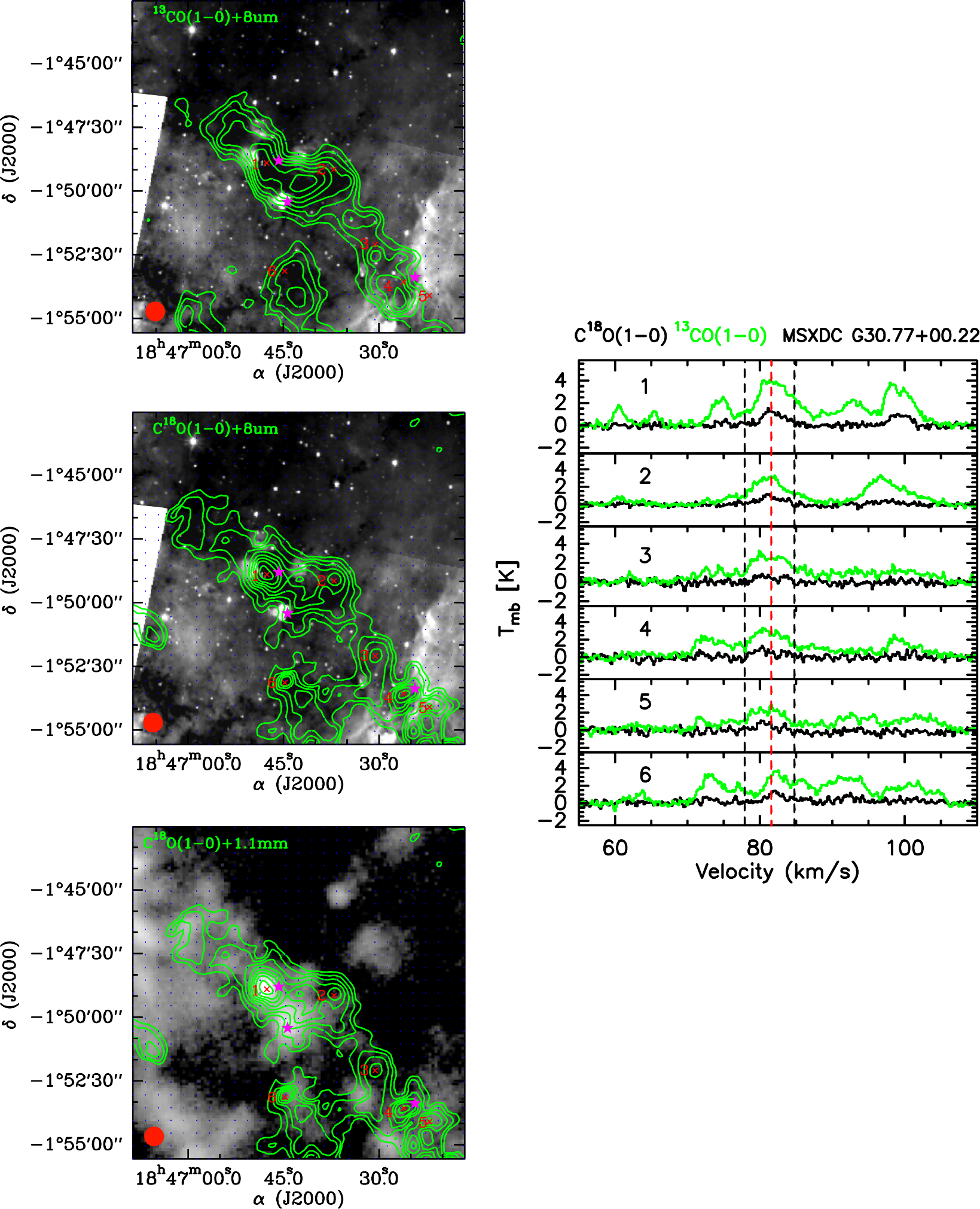}
\vspace{10mm}\caption{Continued}
\end{figure*}

\setcounter{figure}{0}
 \clearpage
\begin{figure*}
\vspace{15mm}
 \centering
\includegraphics[angle=0]{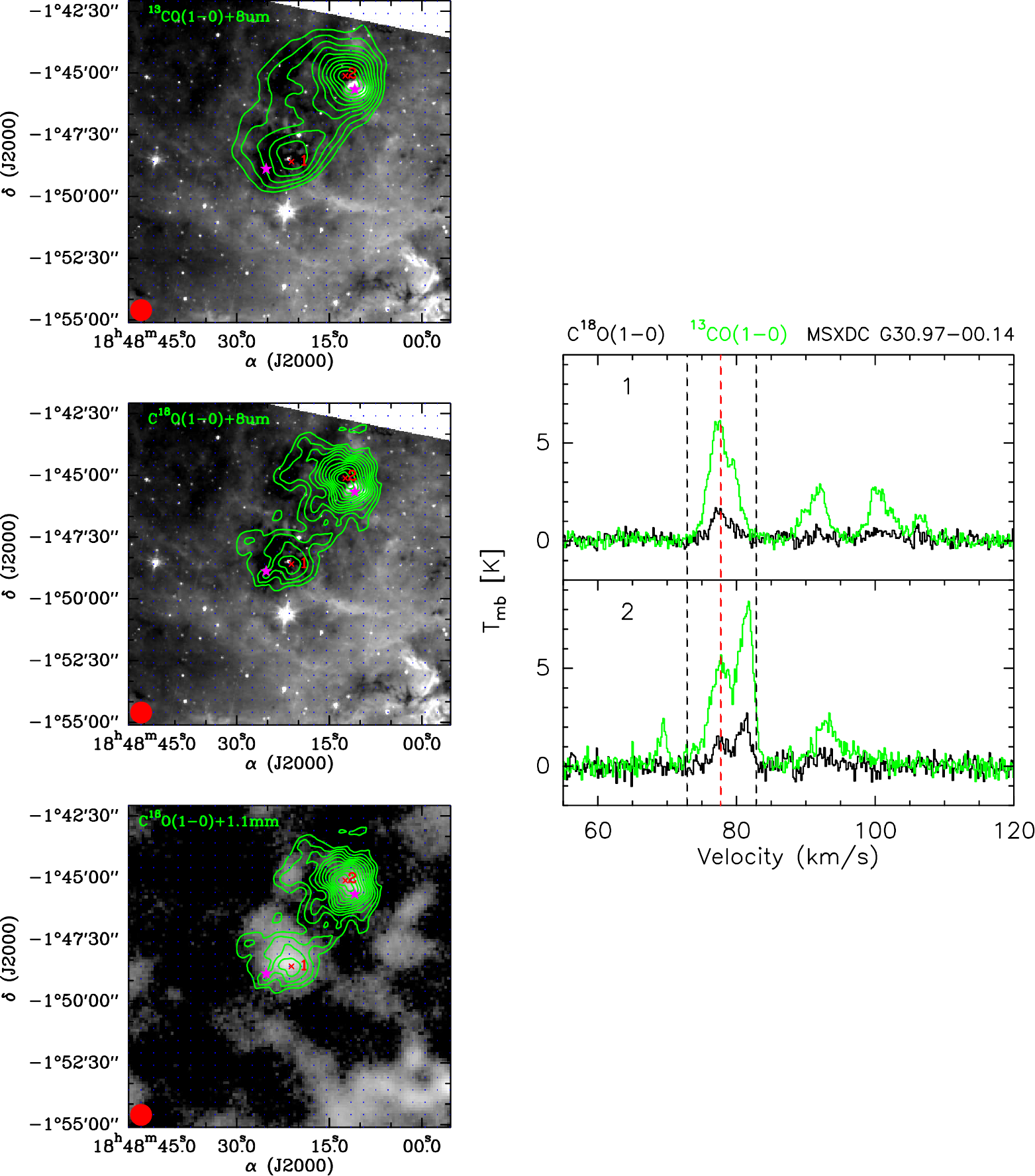}
\vspace{10mm}\caption{Continued}
\end{figure*}

\setcounter{figure}{0}
 \clearpage
\begin{figure*}
\vspace{15mm}
 \centering
\includegraphics[angle=0]{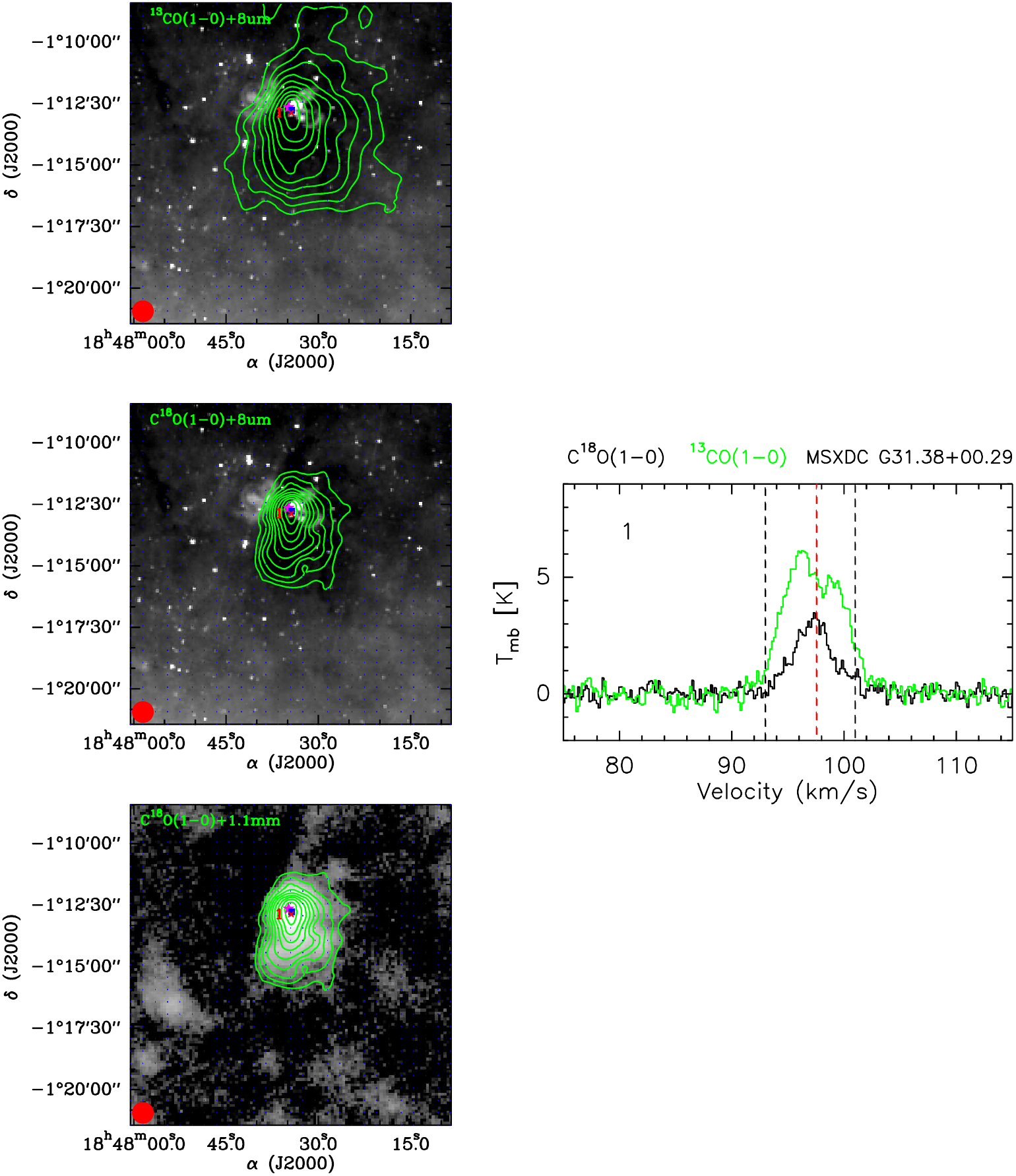}
\vspace{10mm}\caption{Continued}
\end{figure*}

\setcounter{figure}{0}
 \clearpage
\begin{figure*}
\vspace{15mm}
 \centering
\includegraphics[angle=0]{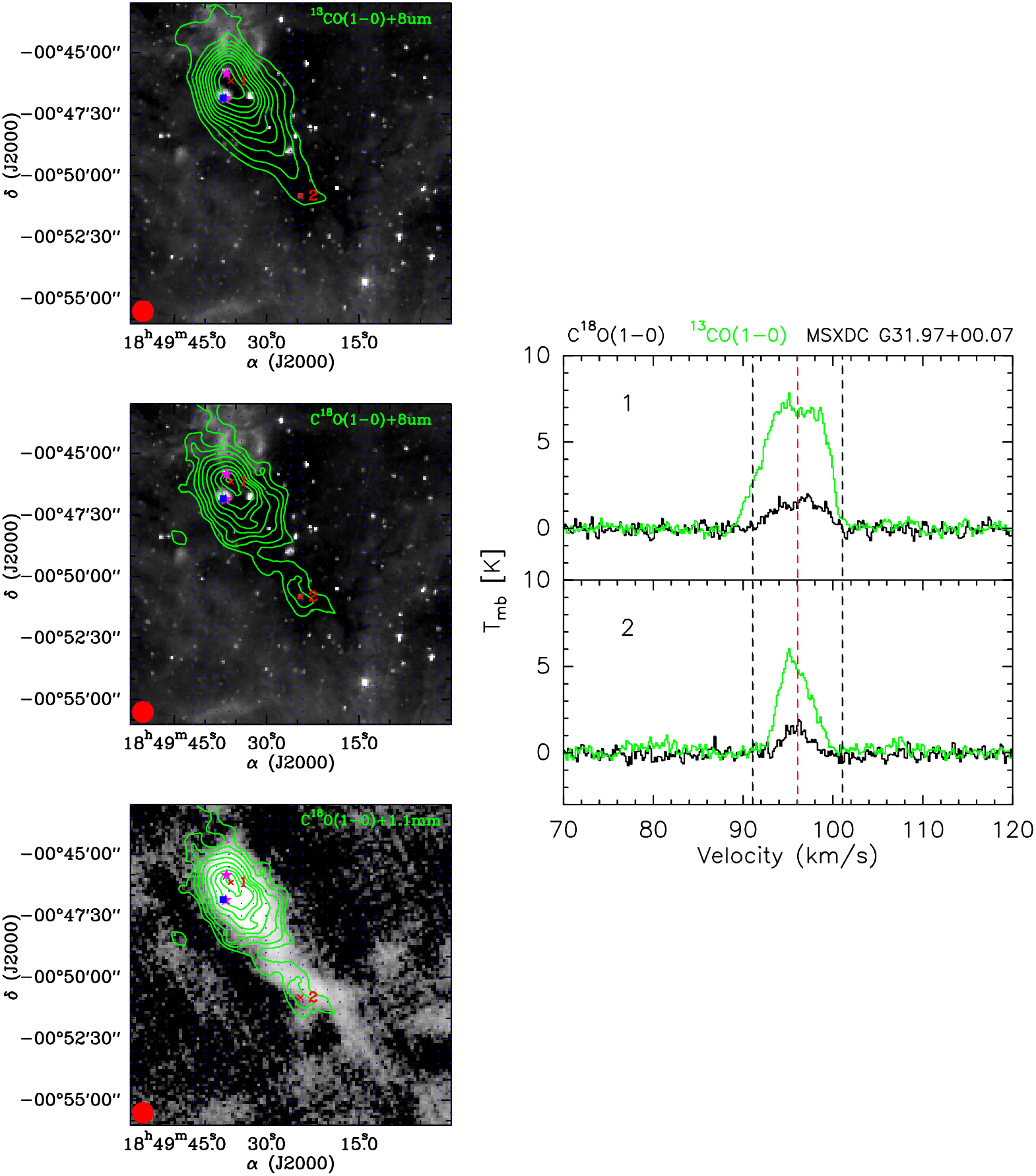}
\vspace{10mm}\caption{Continued}
\end{figure*}

\setcounter{figure}{0}
 \clearpage
\begin{figure*}
\vspace{10mm}
 \centering
\includegraphics[angle=0]{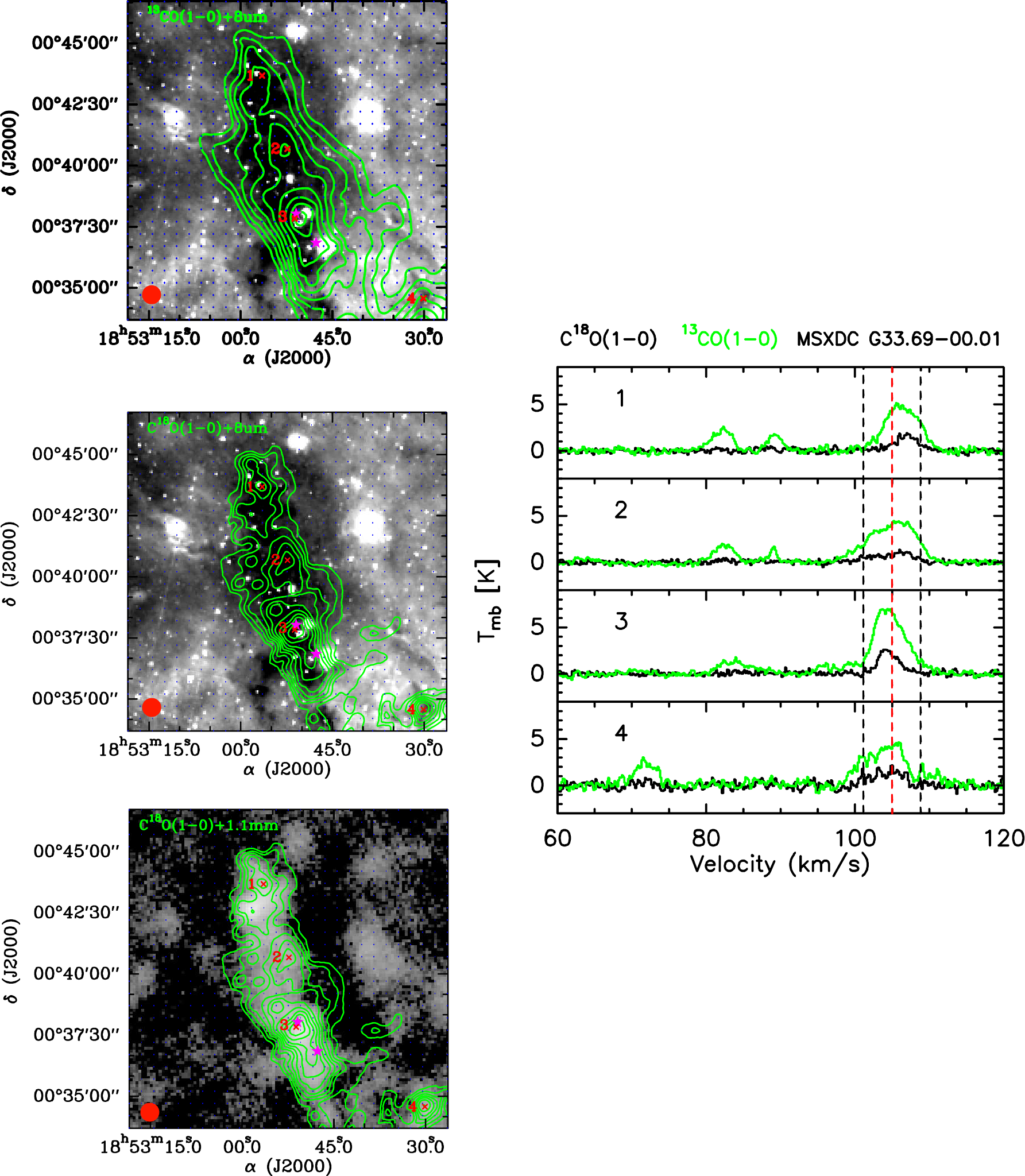}
\vspace{10mm}\caption{Continued}
\end{figure*}

\setcounter{figure}{0}
 \clearpage
\begin{figure*}
\vspace{10mm}
 \centering
\includegraphics[angle=0]{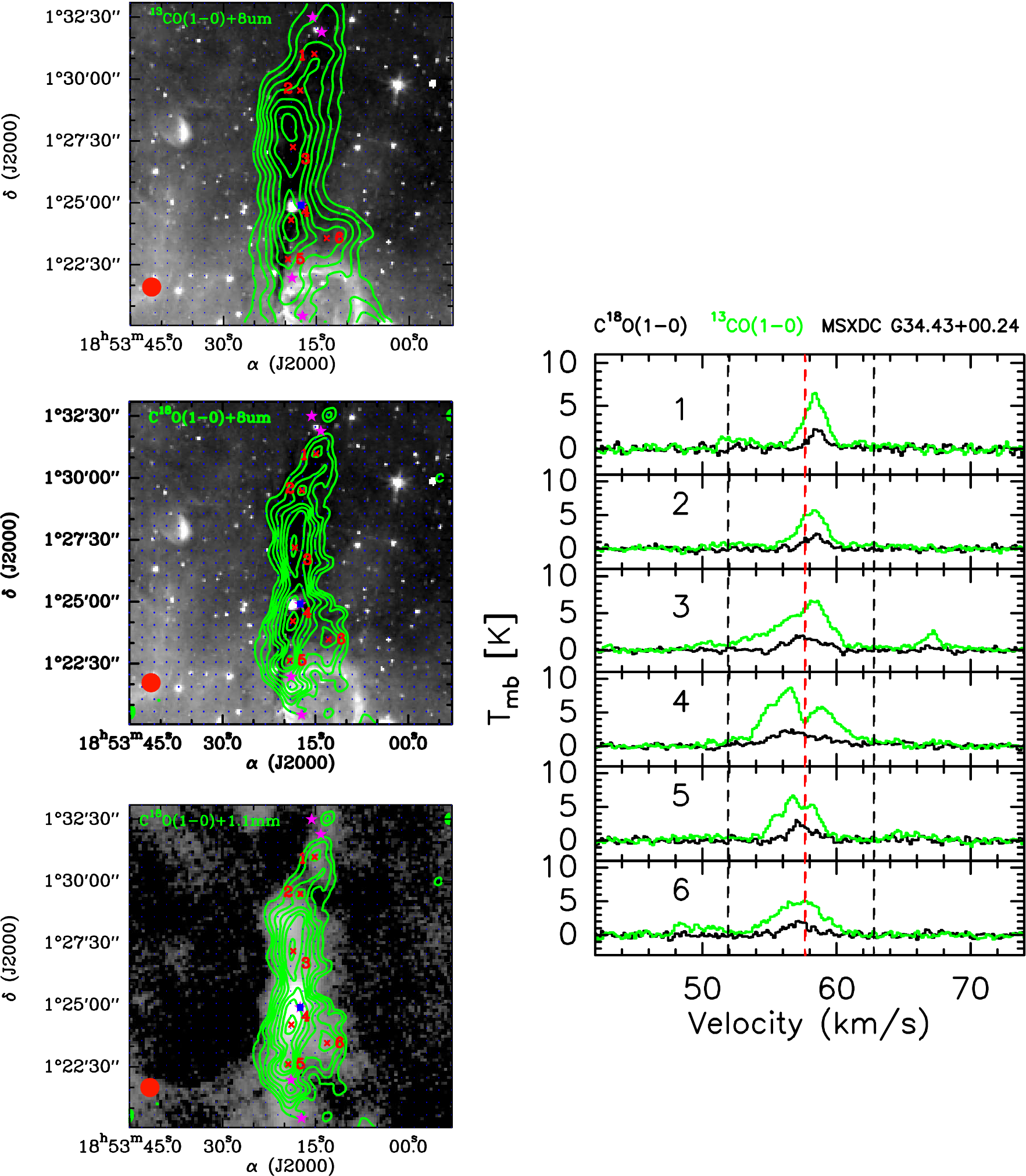}
\vspace{10mm}\caption{Continued}
\end{figure*}

\setcounter{figure}{0}
 \clearpage
\begin{figure*}
\vspace{15mm}
 \centering
\includegraphics[angle=0]{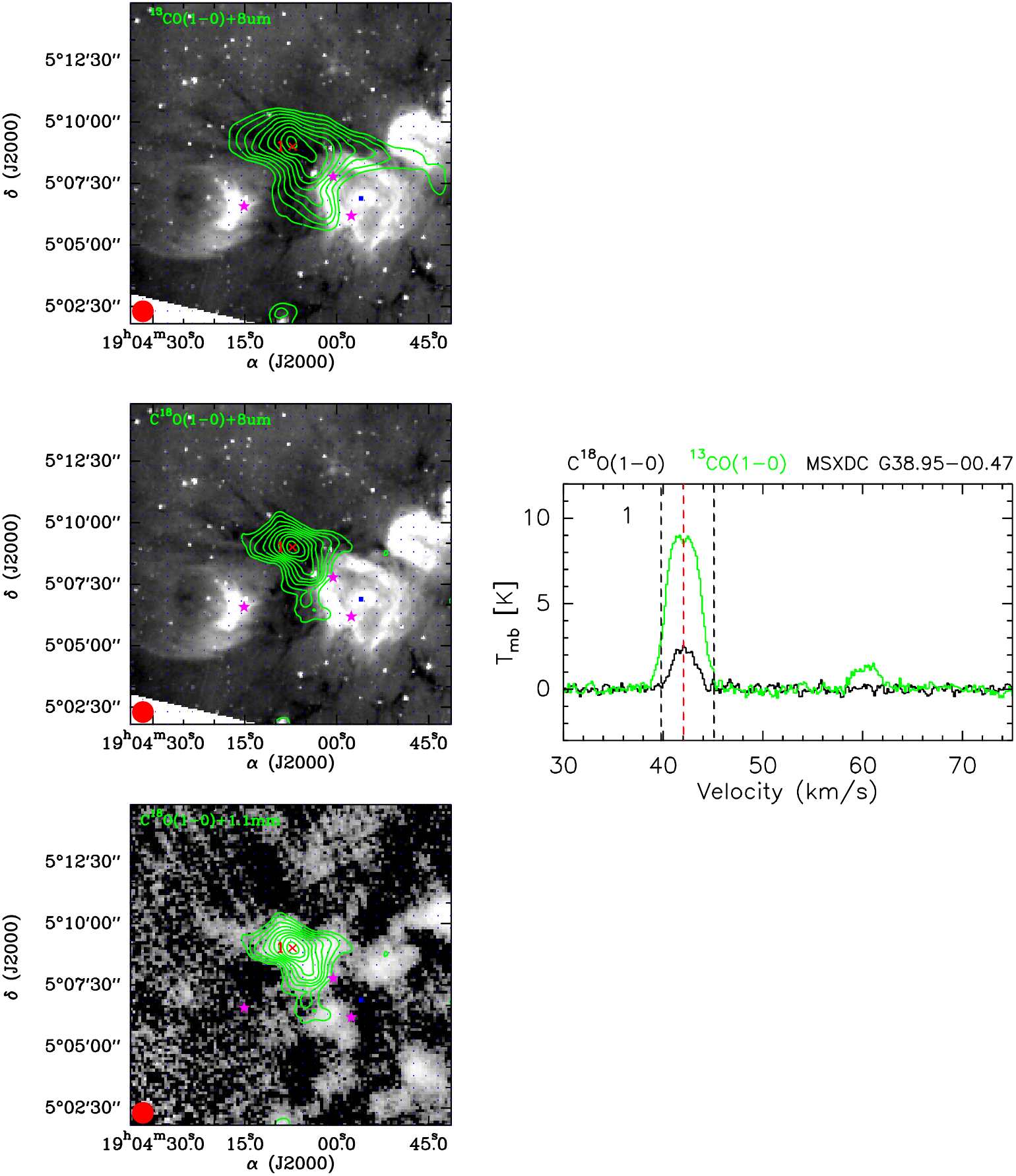}
\vspace{10mm}\caption{Continued}
\end{figure*}

\setcounter{figure}{0}
 \clearpage
\begin{figure*}
\vspace{10mm}
 \centering
\includegraphics[angle=0]{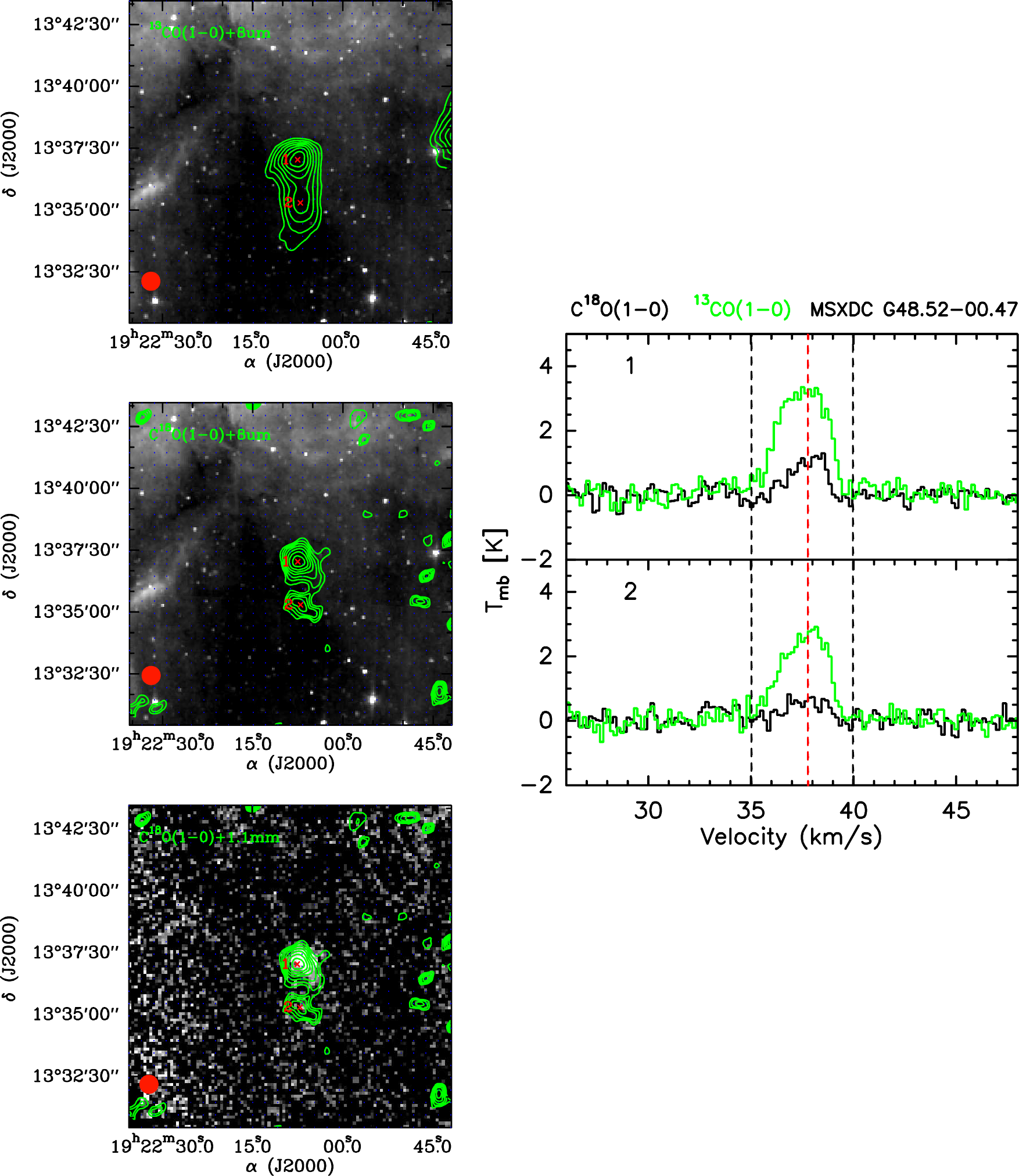}
\vspace{10mm}\caption{Continued}
\end{figure*}

\setcounter{figure}{0}
 \clearpage
\begin{figure*}
\vspace{15mm}
 \centering
\includegraphics[angle=0]{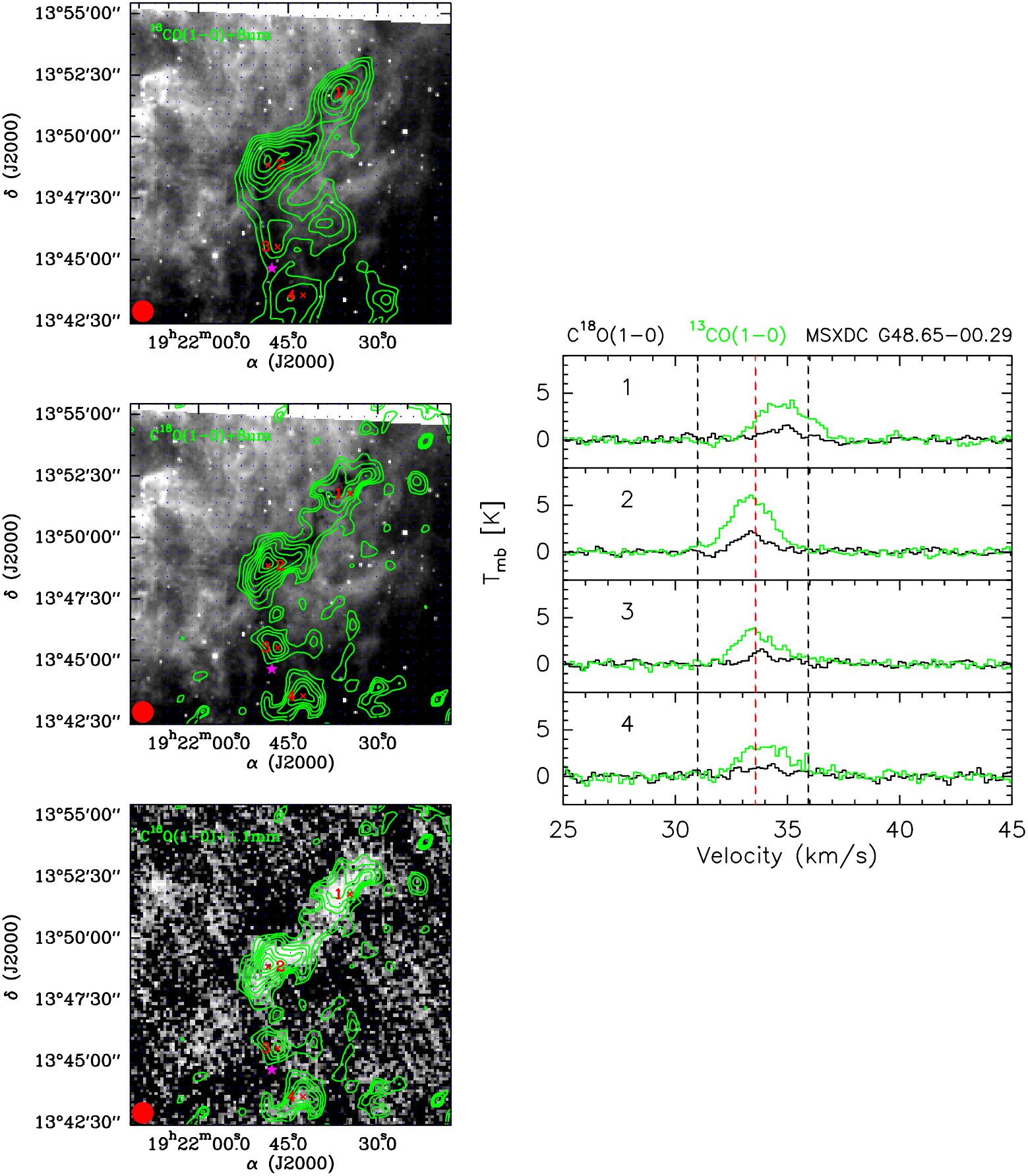}
\vspace{10mm}\caption{Continued}
\end{figure*}

\clearpage
\begin{figure*}
\vspace{5mm}
\begin{minipage}[t]{0.5\linewidth}
  \centering
   \includegraphics[width=80mm,height=50mm]{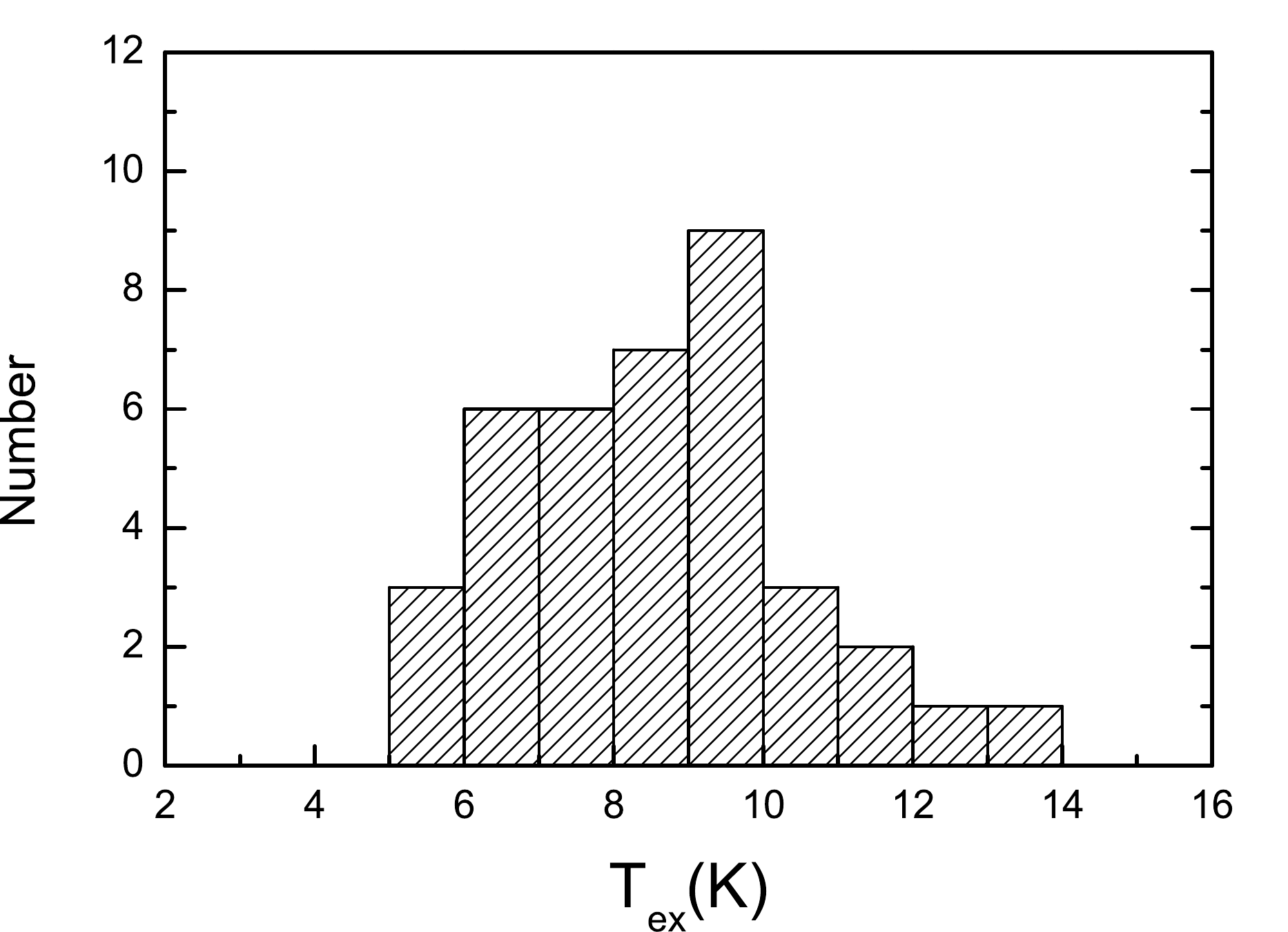}
  \end{minipage}%
  \begin{minipage}[t]{0.5\linewidth}
  \centering
   \includegraphics[width=80mm,height=50mm]{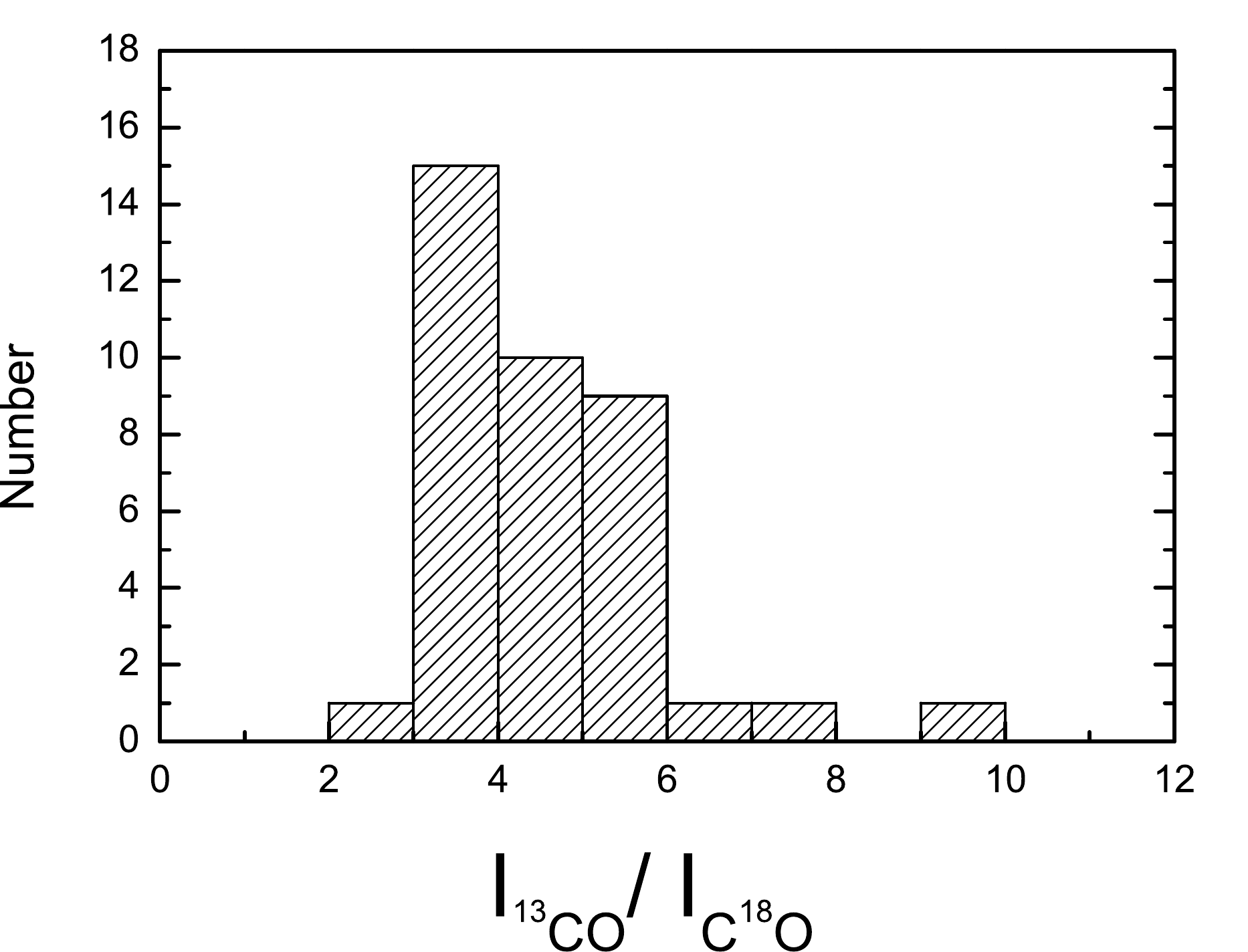}
  \end{minipage}%
\vspace{5mm}
\begin{minipage}[t]{0.5\linewidth}
  \centering
   \includegraphics[width=80mm,height=50mm]{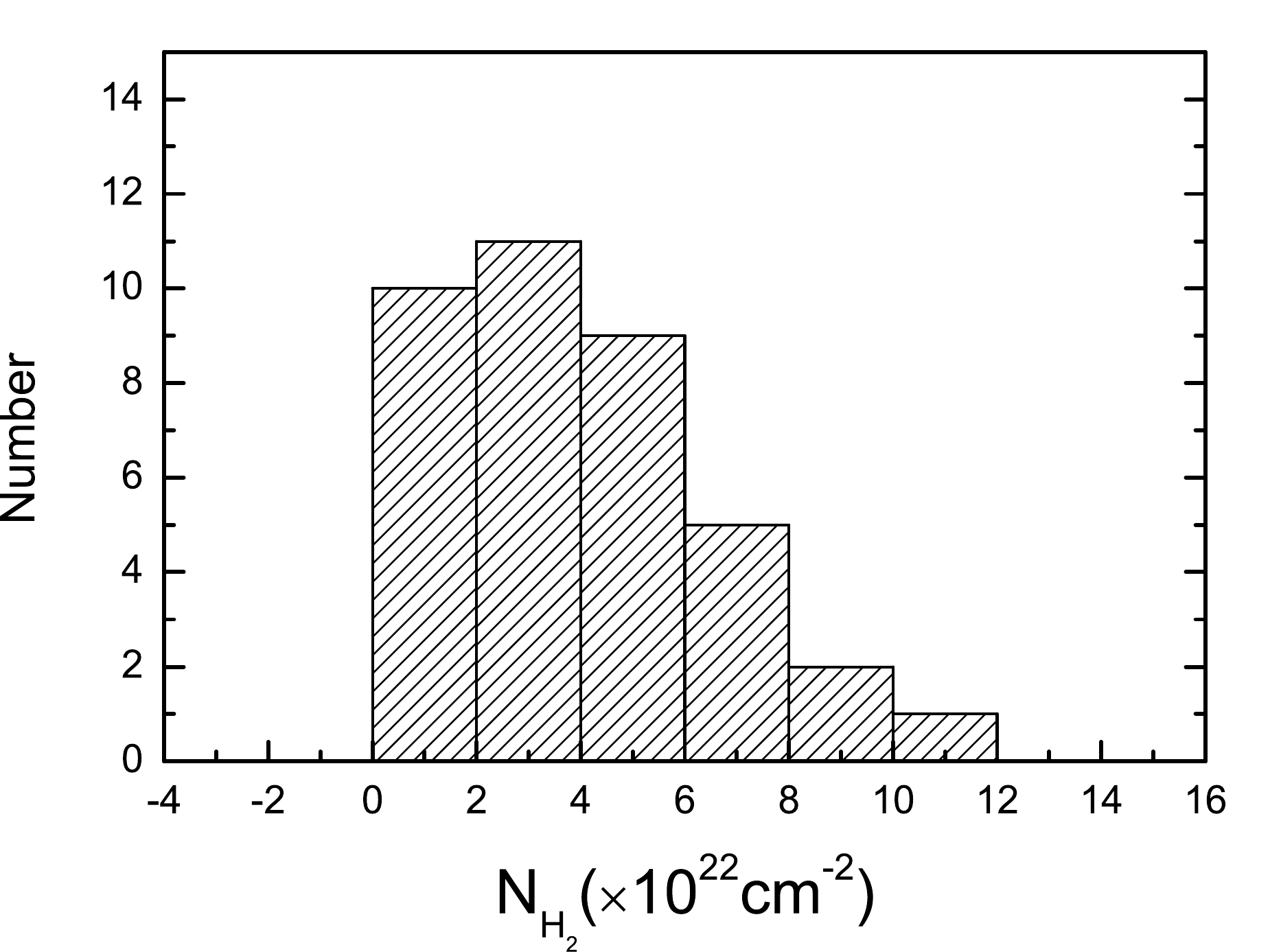}
  \end{minipage}%
  \begin{minipage}[t]{0.5\linewidth}
  \centering
   \includegraphics[width=80mm,height=50mm]{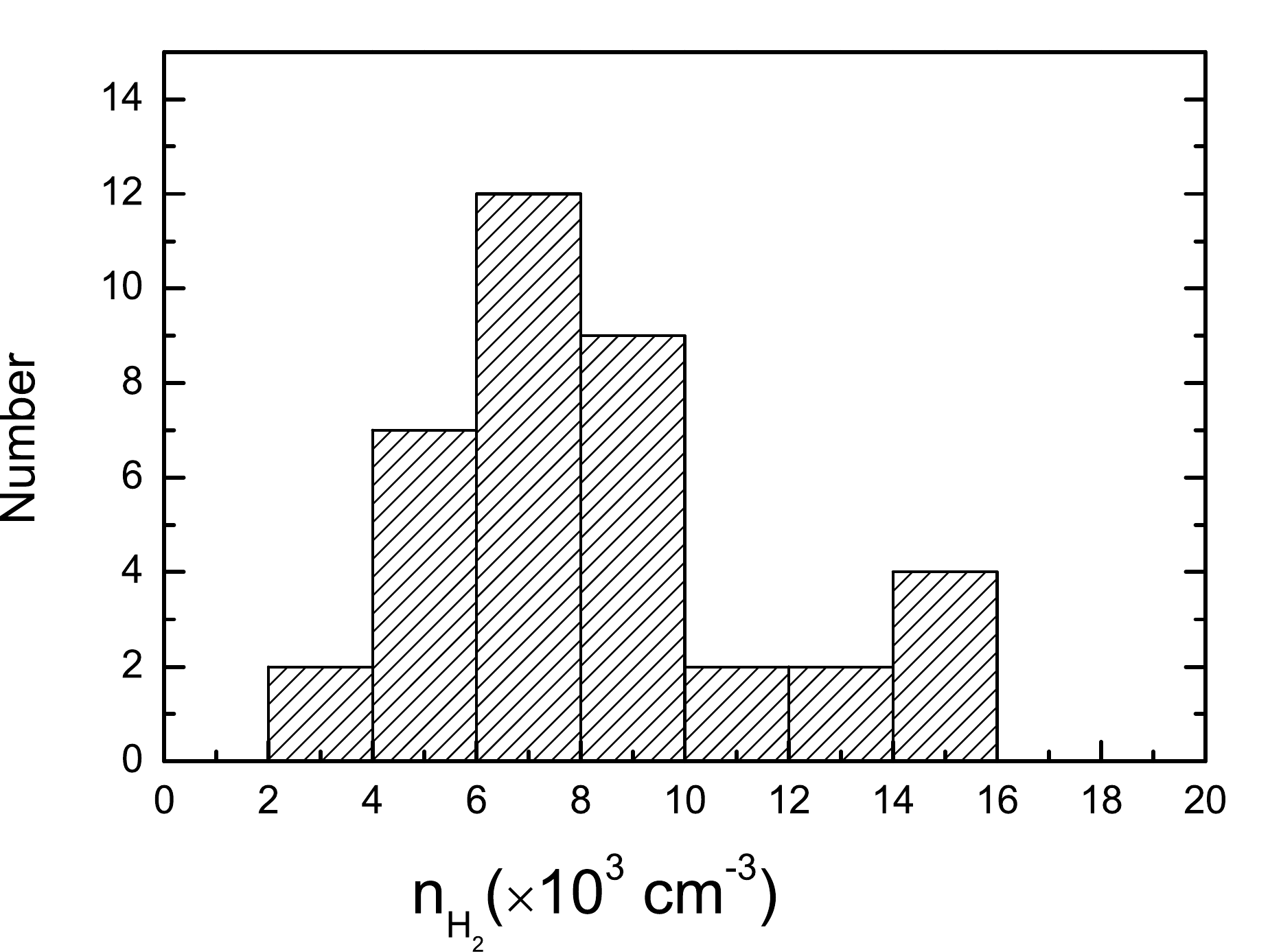}
  \end{minipage}%
  \vspace{5mm}
\begin{minipage}[t]{0.5\linewidth}
  \centering
   \includegraphics[width=80mm,height=50mm]{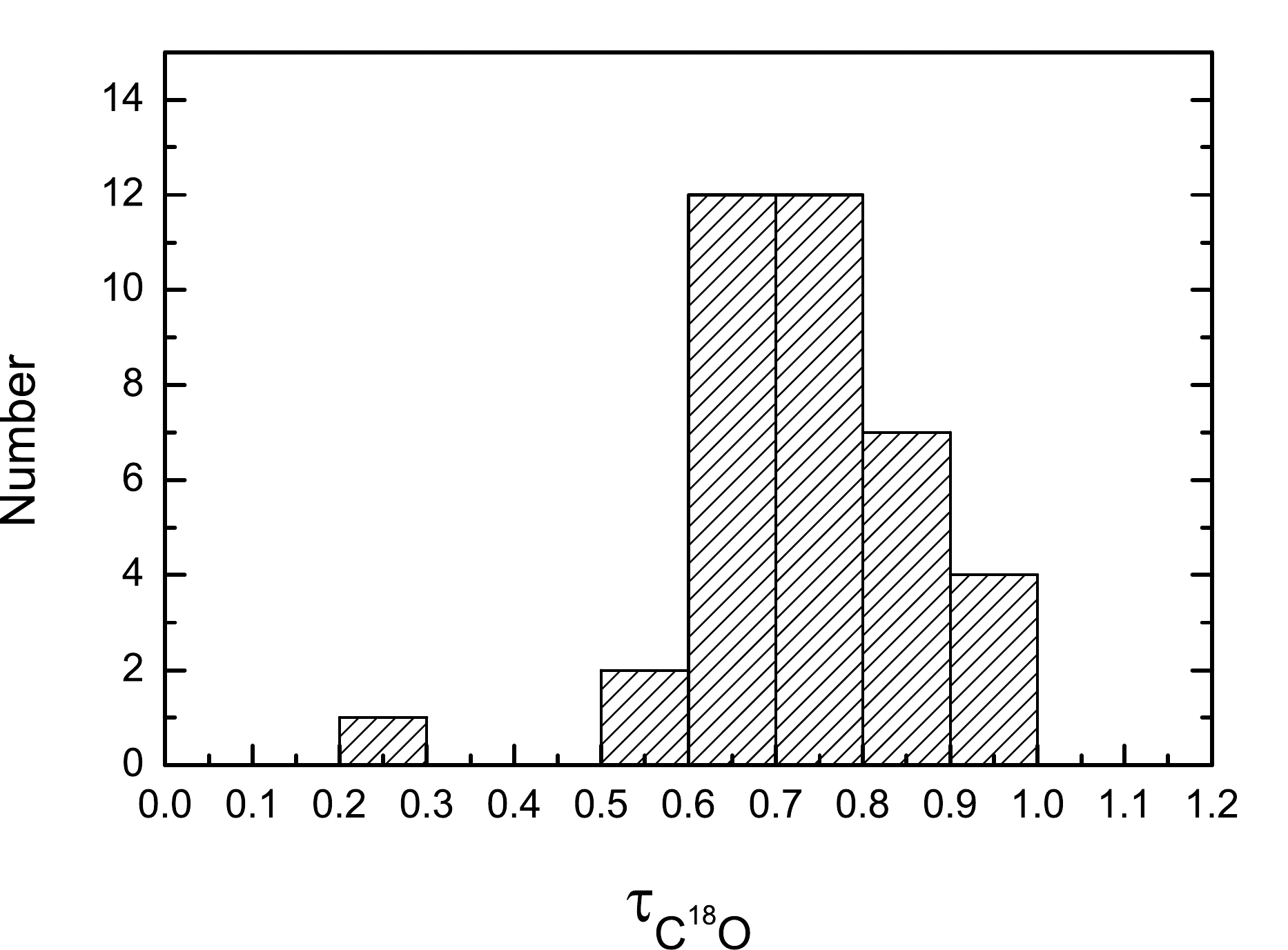}
  \end{minipage}%
  \begin{minipage}[t]{0.5\linewidth}
  \centering
   \includegraphics[width=80mm,height=50mm]{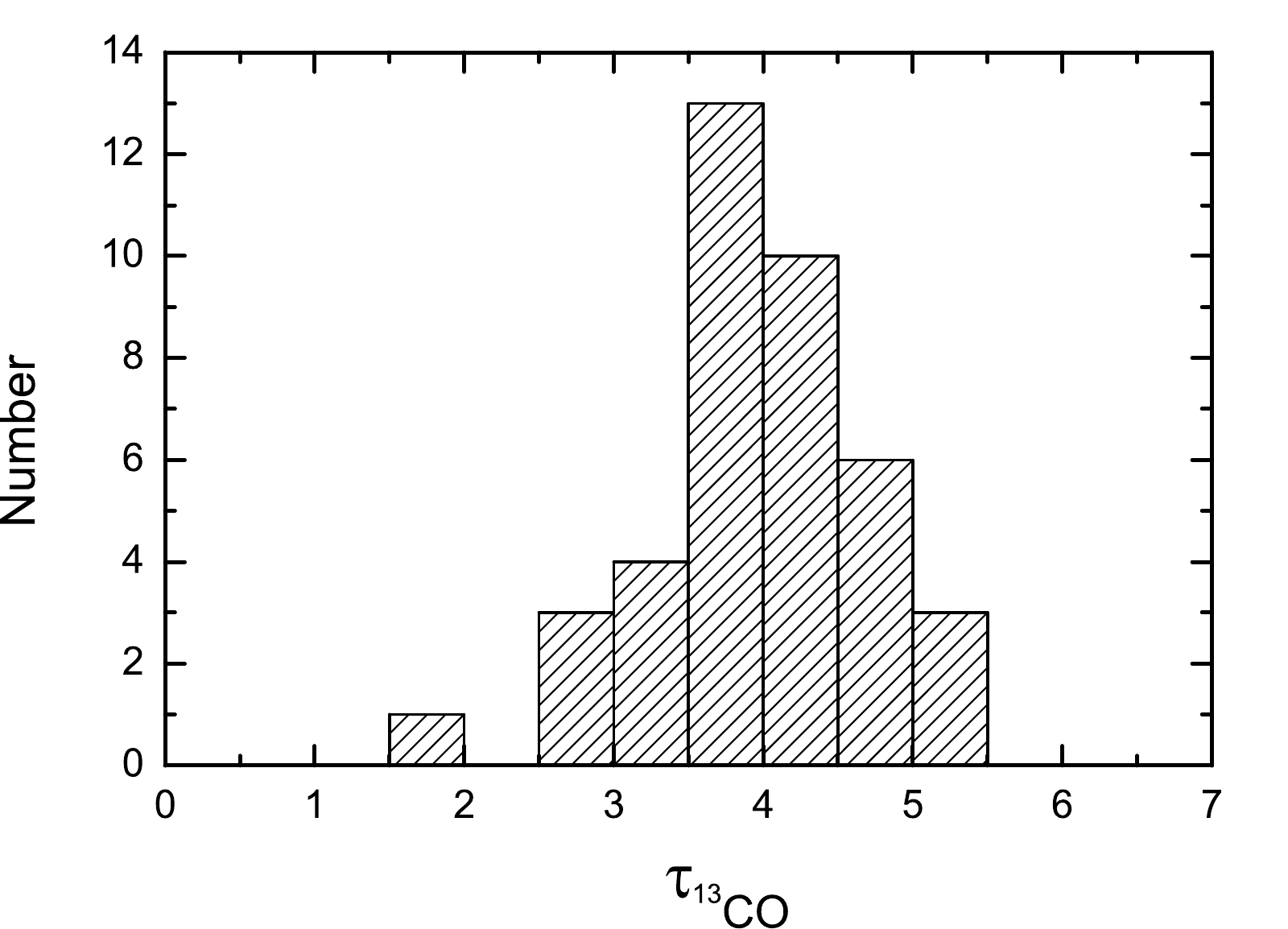}
  \end{minipage}%
   \vspace{5mm}
\begin{minipage}[t]{0.5\linewidth}
  \centering
   \includegraphics[width=80mm,height=50mm]{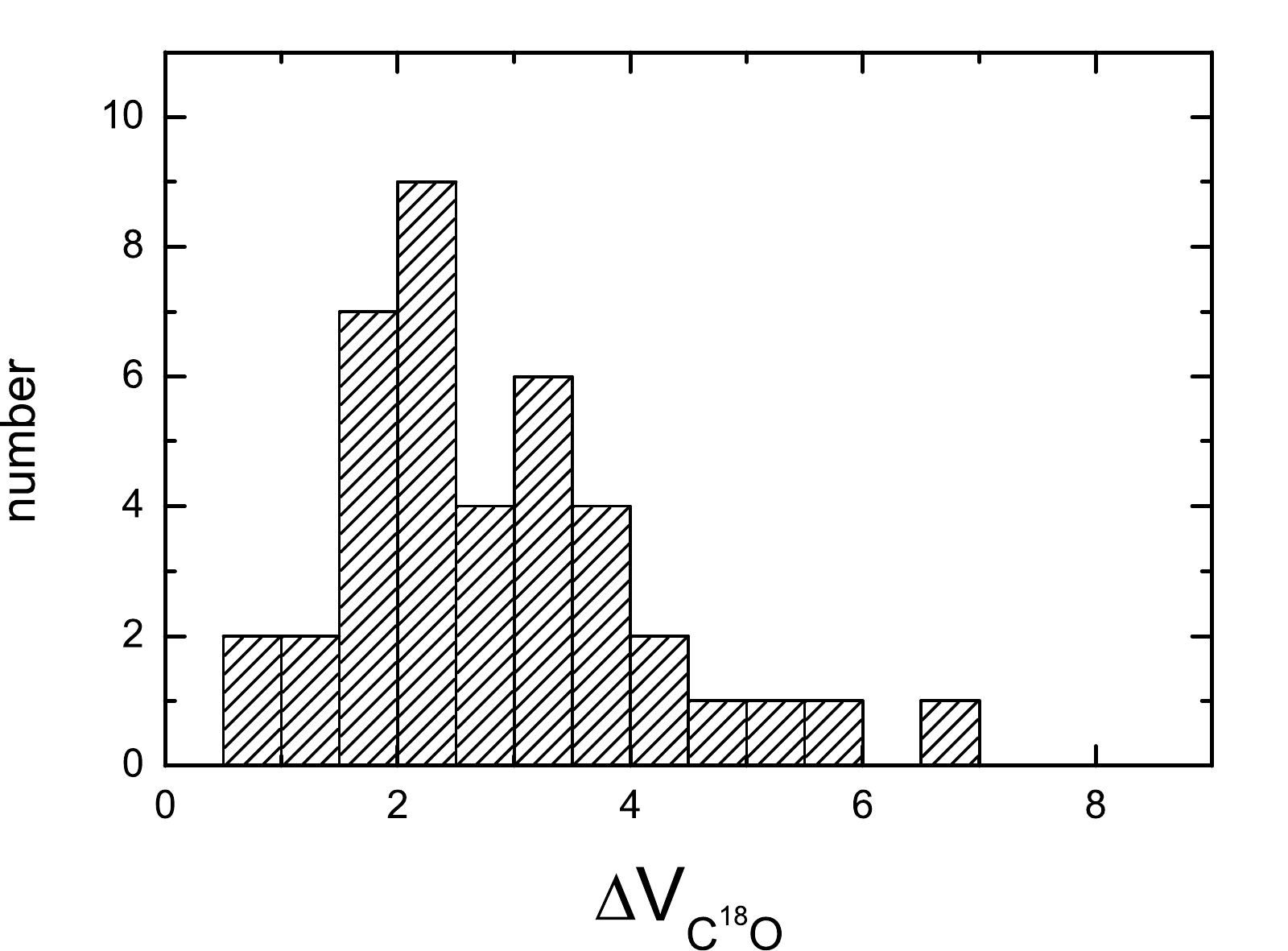}
  \end{minipage}%
  \begin{minipage}[t]{0.5\linewidth}
  \centering
   \includegraphics[width=80mm,height=50mm]{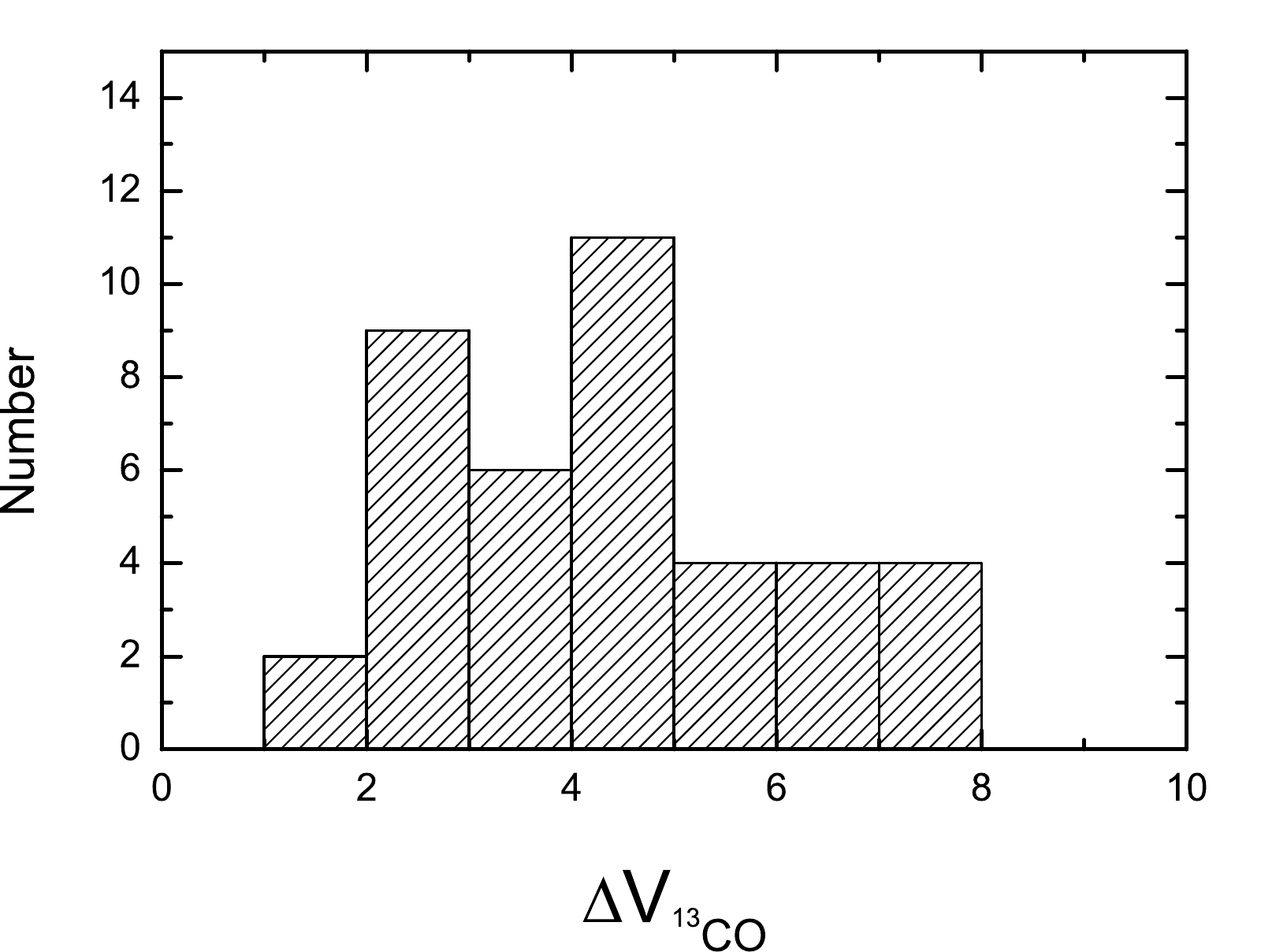}
  \end{minipage}%
\end{figure*}

\begin{figure*}
\vspace{5mm}
  \begin{minipage}[t]{0.5\textwidth}
  \centering
   \includegraphics[width=80mm,height=50mm]{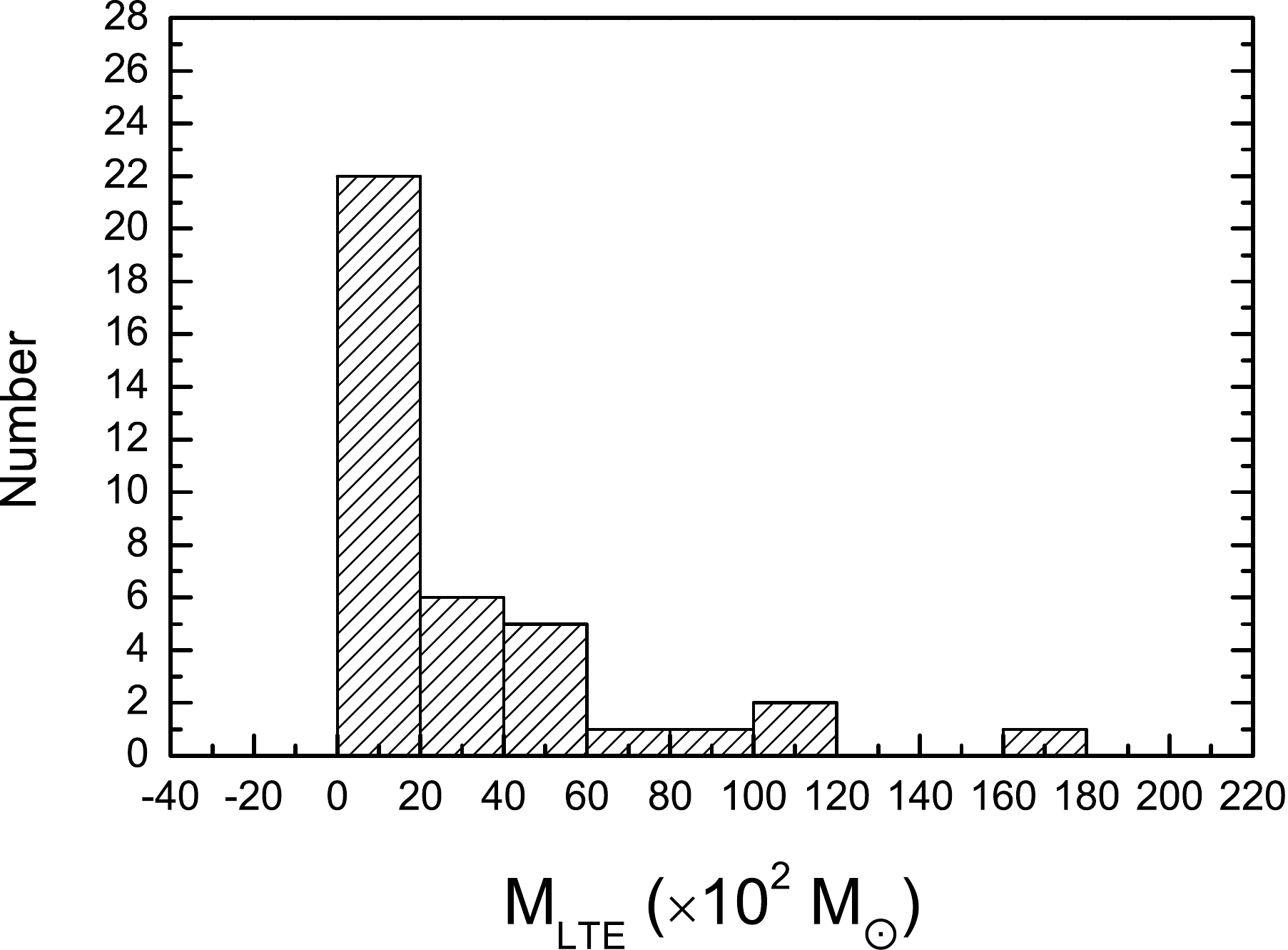}
  \end{minipage}%
  \begin{minipage}[t]{0.5\textwidth}
  \centering
   \includegraphics[width=80mm,height=50mm]{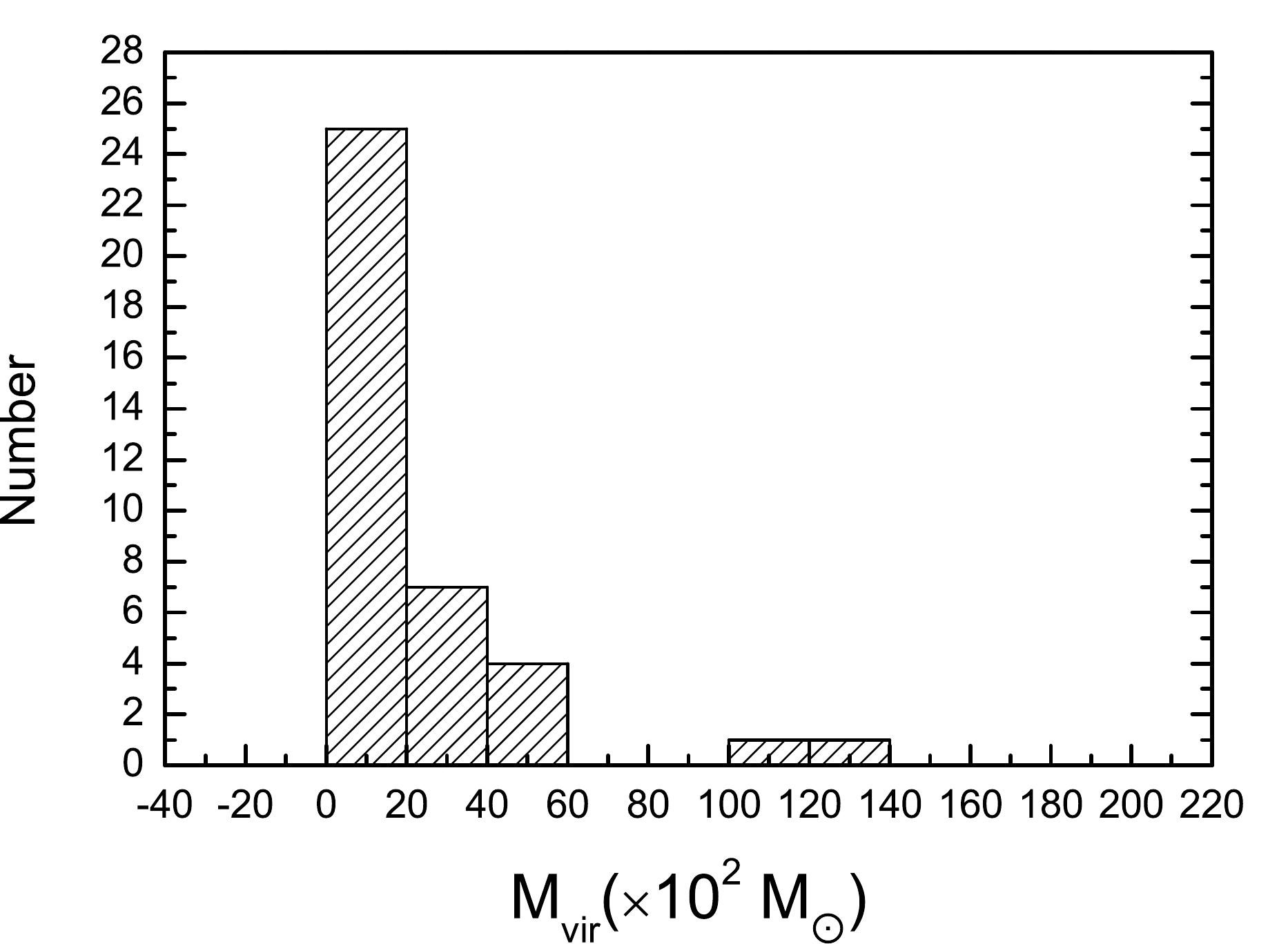}
  \end{minipage}%
\vspace{5mm}
\begin{minipage}[t]{0.5\textwidth}
  \centering
   \includegraphics[width=80mm,height=50mm]{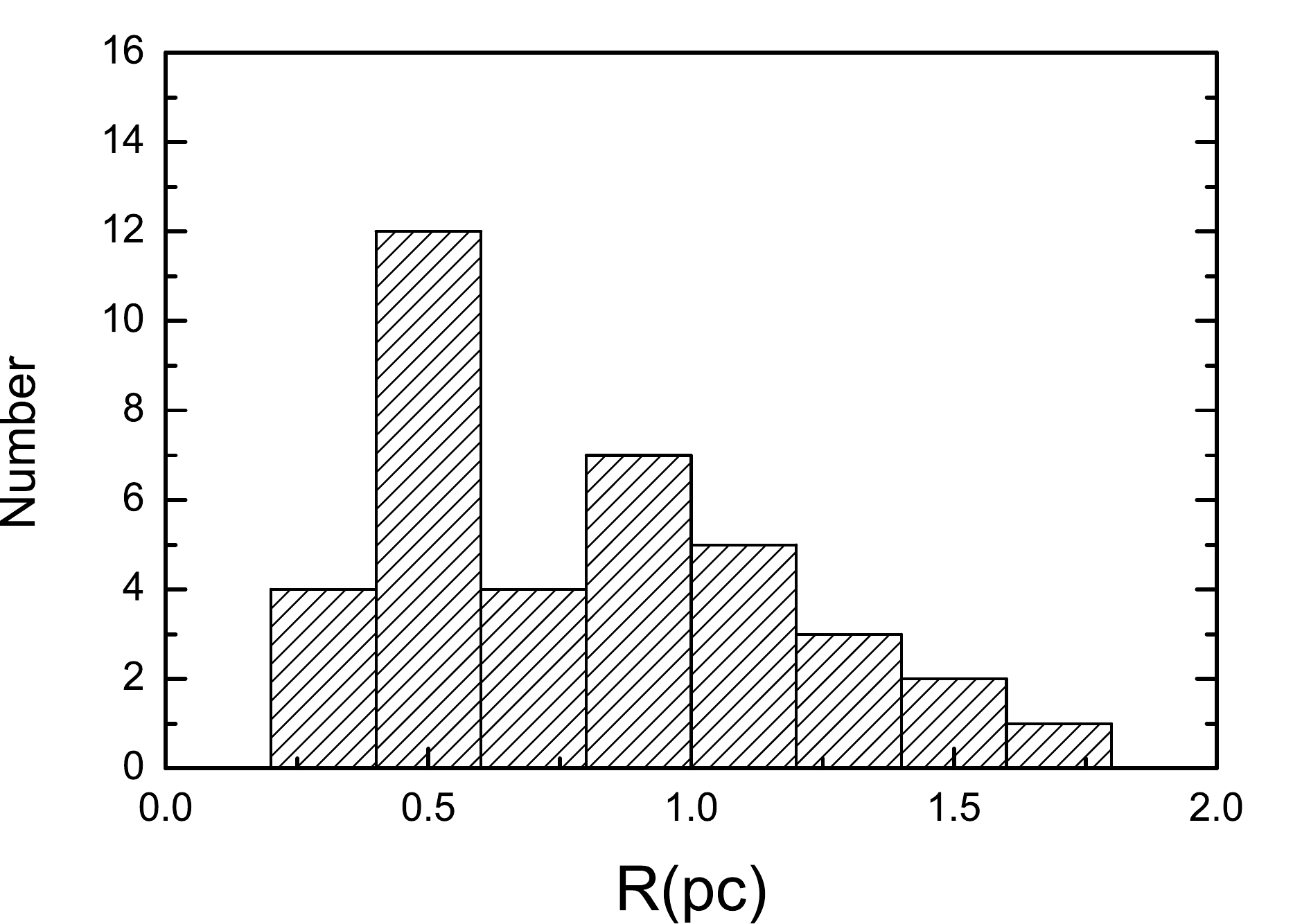}
  \end{minipage}%
\caption{The histograms of the physical parameters towards the 40
IRDC cores.}
\end{figure*}

\begin{figure*}
%\vspace{100mm}
\includegraphics[width=110mm,height=80mm,angle=0]{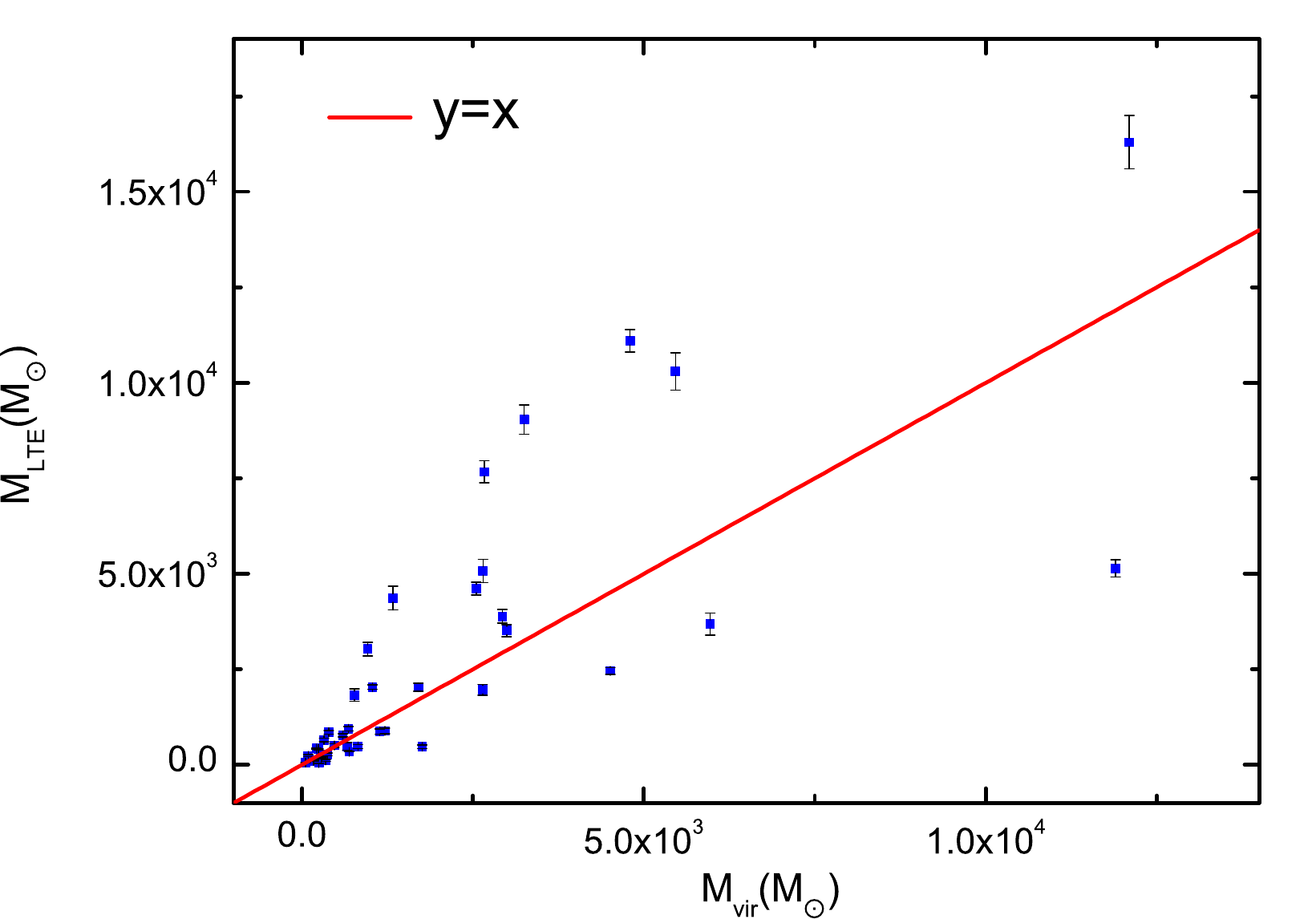}
\caption{The relation graph between $\rm M_{LTE}$ and $\rm M_{vir}$.
The red line represents the relation of $\rm M_{LTE}=M_{vir}$. }
\end{figure*}

\clearpage
\begin{figure*}
\vspace{5mm}
\includegraphics[angle=0]{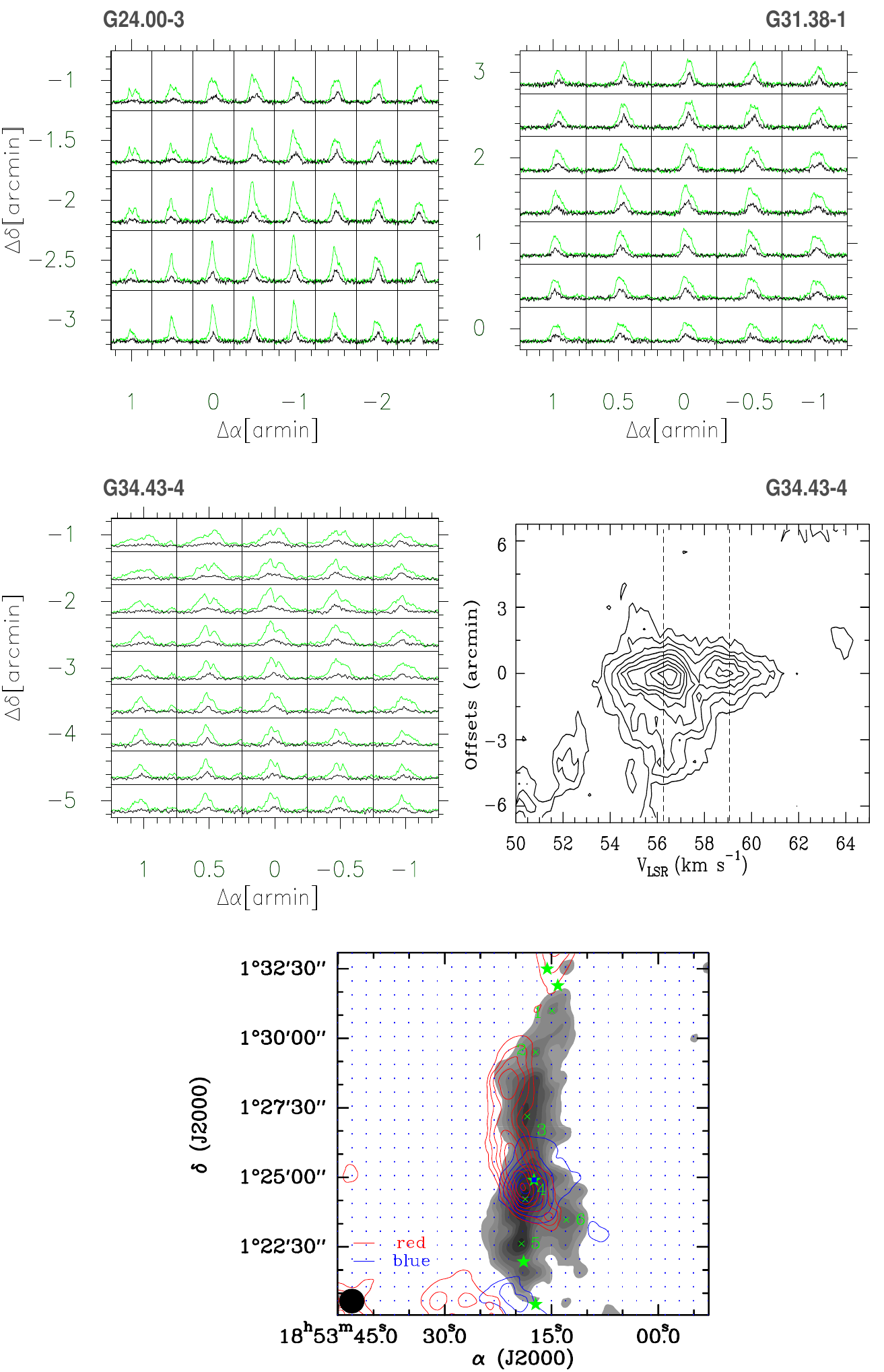}
\vspace{2mm} \caption{Top Left, top right, and middle left: the
mapping grids of IRDC core G24.00-3, G31.38-1 and G34.43-4. The
green lines and the black lines represent the $\rm ^{13}CO$ (1-0)
and $\rm C^{18}O$ (1-0) lines, respectively. Middle right: the
position-velocity diagram of IRDC core G34.43-4 in $\rm ^{13}CO$,
the two vertical lines indicate the beginning of the blue and red
wings, respectively. Bottom: the outflow contours overlaid on the
$\rm C^{18}O$ (1-0) integrated intensity map (Grey) of IRDC core
G34.43-4 in $\rm ^{13}CO$ (1-0) line, the integrated ranges in the
blue and red wings are $\rm 52\,km s^{-1}\sim56.5 \,km s^{-1}$ and
$\rm 59.2\,km s^{-1}\sim62.2 \,km s^{-1}$, respectively. The contour
levels are $40\% \sim 100\%$ of each wing¨s peak value. The green
"$\times$" and corresponding numbers mark the centers of the cores.
The green $\star$ and the blue box represent the IRAS sources and an
UCHII region.}
\end{figure*}

\begin{figure*}
\vspace{10mm}
  \begin{minipage}[t]{0.5\textwidth}
  \centering
   \includegraphics[width=85mm,height=65mm]{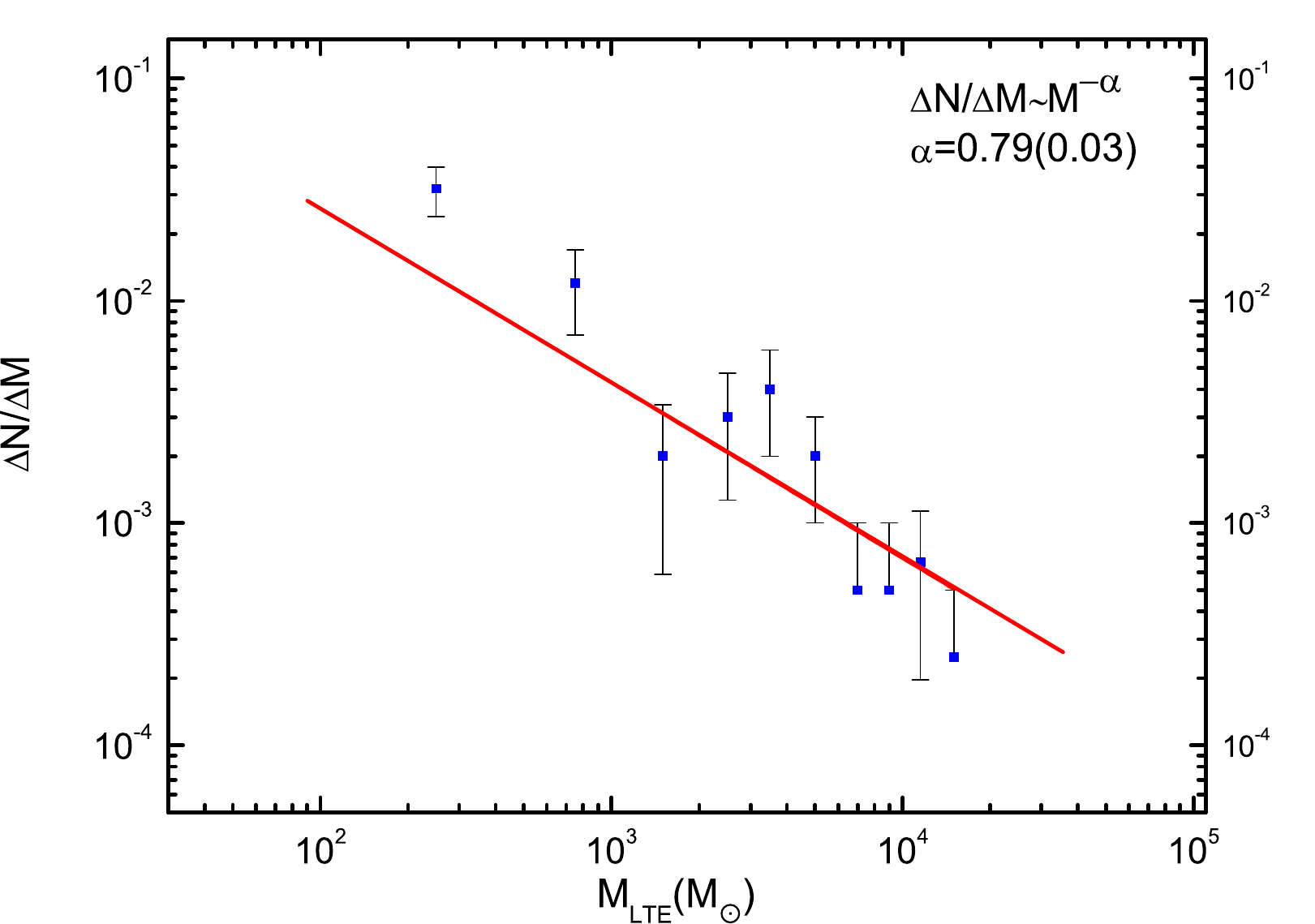}
  \end{minipage}%
  \begin{minipage}[t]{0.5\textwidth}
  \centering
   \includegraphics[width=85mm,height=65mm]{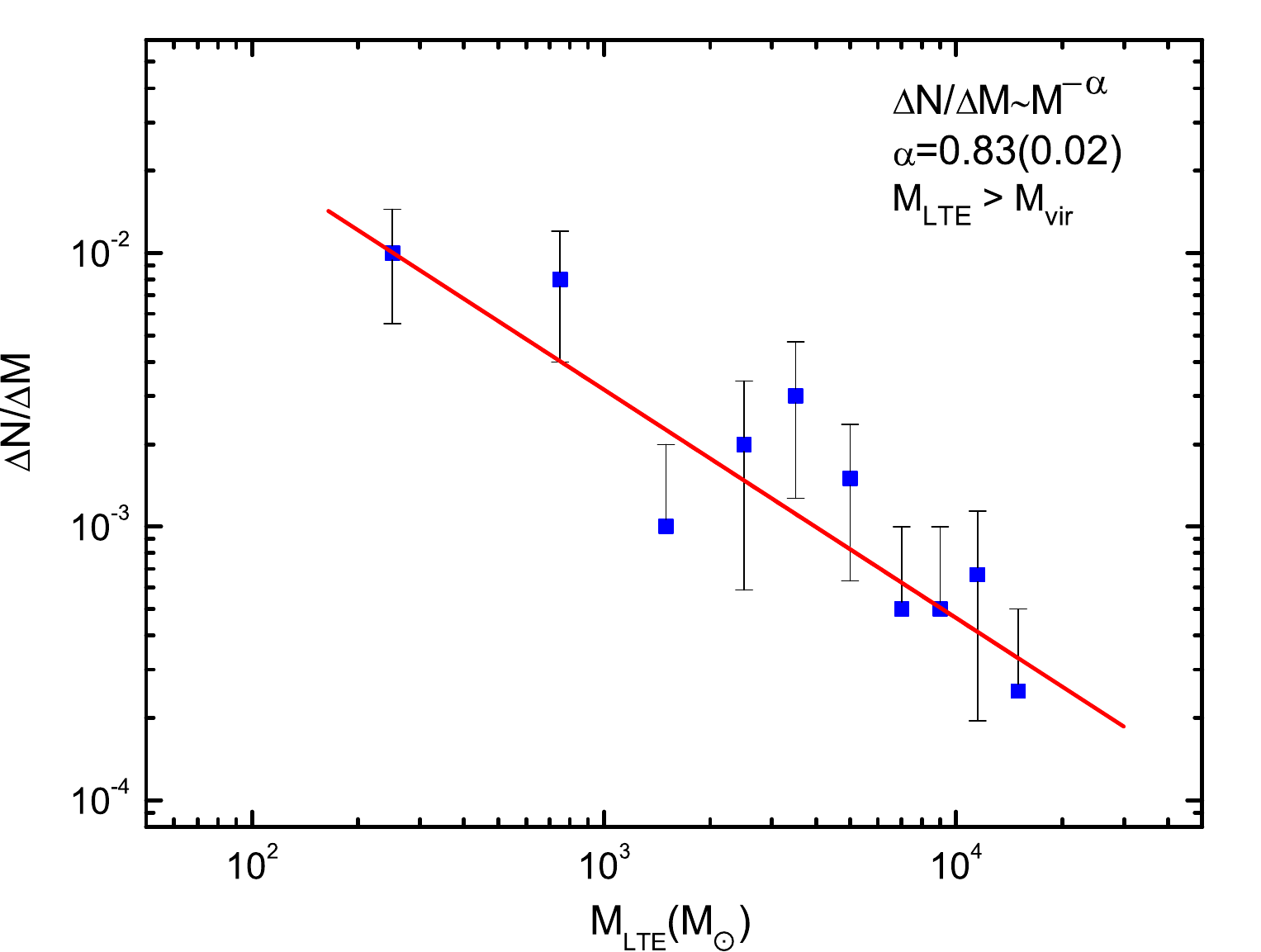}
  \end{minipage}%
\caption{Left: the mass spectrum of the whole IRDC cores. The best
fit to the data is a power-law function $\rm\Delta N/\Delta M\sim
M^{-\alpha}$ with $\alpha=0.79\pm0.03$. Right: the mass spectrum of
the IRDC cores with $\rm M_{LTE}>M_{vir}$  and has a power-law index
$0.83\pm0.02$. The axes are in logarithmic units.}
\end{figure*}

\end{document}